\documentclass[aps,prb,twocolumn,amsmath,amssymb,superscriptaddress,floatfix]{revtex4-1} 

\usepackage{amsmath,amsthm,amssymb}
\usepackage[dvipdfmx]{graphicx}
\usepackage{amsmath,bm}
\usepackage{color,ulem}
\newcommand{\1}{\mbox{1}\hspace{-0.25em}\mbox{l}}

\let\olditemize\itemize
\renewcommand{\itemize}{
   \olditemize
   \setlength{\itemsep}{1pt}
   \setlength{\parskip}{0pt}
   \setlength{\parsep}{0pt}
}

\newlength{\figwidth}
\newlength{\figlarge}
\begin{document}
\title{Correlation effects on topological crystalline insulators}
\author{Tsuneya Yoshida} 
\affiliation{Condensed Matter Theory Laboratory, RIKEN, Wako, Saitama, 351-0198, Japan}
\author{Akira Furusaki} 
\affiliation{Condensed Matter Theory Laboratory, RIKEN, Wako, Saitama, 351-0198, Japan}
\affiliation{RIKEN Center for Emergent Matter Science (CEMS), Wako, Saitama, 351-0198, Japan}
\date{\today}
\begin{abstract}
We study interaction effects on the topological crystalline insulators
protected by time-reversal ($T$) and reflection symmetry ($R$)
in two and three spatial dimensions.
From the stability analysis of the edge states with bosonization,
we find that the classification of the two-dimensional SPT phases protected by
$Z_2\times[\mbox{U(1)}\rtimes T]$ symmetry is reduced from $\mathbb{Z}$
to $\mathbb{Z}_4$ by interactions,
where the $Z_2$ symmetry denotes the reflection whose mirror plane
is the two-dimensional plane itself.
By extending the approach recently proposed by Isobe and Fu, we show
that the classification of
the three-dimensional SPT phases (i.e., topological crystalline insulators)
protected by $R\times[\mbox{U(1)}\rtimes T]$ symmetry is reduced
from $\mathbb{Z}$ to $\mathbb{Z}_8$ by interactions.
\end{abstract}
\pacs{
03.65.Vf, 
71.27.+a, 
} 
\maketitle


\section{Introduction}
Recently, topological structures of gapped quantum states
have attracted much attention.
The topologically nontrivial phases are many-body states with gapped
excitation spectra in the bulk and characterized by nontrivial
topological structure of wave functions.
A remarkable property of topological phases is the existence of stable
gapless boundary modes, which is a source of various exotic properties;
the boundary modes are
the origin of quantization of the Hall conductivity
in integer quantum Hall systems\cite{TKNN}
and topological magnetoelectric effects in three-dimensional
topological insulators.\cite{Qi_3DTI2008}
One of the important questions in this field is classification of
topological phases, i.e., to count how many topologically distinct phases
exist under given symmetry.
The first answer to this question is obtained for free-fermion systems,
for which classification is summarized in the so-called periodic table of
topological insulators and superconductors.\cite{Schnyder_classification_free_2008,Ryu_classification_free_2010,Kitaev_classification_free_2009}
Free-fermion systems are categorized into the
ten Altland-Zirnbauer symmetry classes\cite{Altland}
in terms of time-reversal, particle-hole and sublattice symmetry.
The periodic table tells us that, in every spatial dimension,
three and two out of the ten Altland-Zirnbauer symmetry classes
have topological insulators/superconductors characterized
by topological indices from $\mathbb{Z}$ and $\mathbb{Z}_2$,
respectively.
The periodic table of topological insulators provides helpful information
for searching for topological materials.

The notion of topological insulators and superconductors is extended
to systems of interacting fermions or bosons.
The symmetry-protected topological (SPT) phases\cite{chen_cohomology_3D}
are many-body ground states with short-range entanglement
which have gapped excitations in the bulk and
gapless excitations on the boundary which are stable
against any symmetry-preserving perturbation.
The Haldane gap phase of integer-spin chains\cite{Chen_classification_1D_1,Pollmann2012} and
bosonic integer quantum Hall states\cite{Senthil_IQHE_2013,furukawa_IQHE_2013}
are examples of bosonic SPT phases in one and two dimensions.
As for fermionic SPT phases,
topologically nontrivial phases are also expected in several $4f$- or
$5d$- electron systems, and the Kondo insulator $\mathrm{SmB_6}$ is
a candidate of a topological insulator of correlated electrons.\cite{Takimoto_2011,Dzero2012,SmB6_Zhang2013,SmB6_Lu2013} 
However, it was shown by Fidkowski and Kitaev that interactions can change
the topological classification (periodic table) of free-fermion
systems.\cite{Z_to_Zn_Fidkowski_10}
They demonstrated that inter-chain interactions in eight Kitaev chains
gap out all Majorana zeromodes without symmetry breaking.
This implies that topological classification of one-dimensional
topological insulators/superconductors in class BDI
is changed from $\mathbb{Z}$ to $\mathbb{Z}_8$ by interactions.
Such collapse of topological classification is also observed
in two dimensions\cite{RyuZhang2012,YaoRyu2013,QiNJP2013,GuLevin2014}
and three dimensions.\cite{Fidkowski_Z162013,Wang_Potter_Senthil2014,Metlitski_3Dinteraction2014,You_Cenke2014}
Several theoretical frameworks for classification of SPT phases
in interacting systems have been developed.
For example, theory based on group cohomology\cite{Chen_classification_1D_1,chen_cohomology_3D,gu_supercohomology}
or cobordisms\cite{kapustin_bosonic_cobordisms2014_1,kapustin_bosonic_cobordisms2014_2,kapustin_fermionic_cobordisms2014}
classifies possible topological actions for the bulk states,
while theoretical approach using the Chern-Simons
theory\cite{Lu_CS_2011,Hsieh_CS_CPT_2014}
or nonlinear sigma models\cite{Fidkowski_Z162013,Wang_Potter_Senthil2014,Metlitski_3Dinteraction2014,Wang_Senthil2014,You_Cenke2014}
examine the stability of gapless boundary modes.

In many cases SPT phases are protected by local symmetry such as time reversal
and other internal symmetries.
However, spatial symmetry (e.g., reflection, rotation etc.) can also
protect topological phases, as pointed out by Fu\cite{Fu_TCI_2011}
and realized in the topological crystalline insulator $\mathrm{SnTe}$.\cite{Tanaka_TCI_SnTe2012,Hsieh_TCH_SnTe_2012} 
This compound respects the time-reversal and reflection symmetry
and has four Dirac cones on the $(0,0,1)$ surface.\cite{Hsieh_TCH_SnTe_2012}
Having a trivial strong $Z_2$ index, $\mathrm{SnTe}$ is a trivial band
insulator as the four gapless Dirac cones are not protected from opening
of a gap by time-reversal symmetry. 
However, the reflection symmetry about the $(1,1,0)$ plane prohibits
any Dirac mass term that could gap out Dirac cones, and the stability
of surface Dirac cones is guaranteed by mirror Chern numbers defined on
mirror planes in the Brillouin zone; the topological classification
is thus $\mathbb{Z}$.\cite{Hsieh_TCH_SnTe_2012,Chiu_TCI_classification2013,Morimoto_classification2013,shiozaki_classification_2014} 
Besides SnTe,
there are other candidate materials theoretically proposed as
topological crystalline insulators
in strongly correlated electron systems:
a heavy-fermion compound\cite{Weng_Dai_TCIinYbB12_2014} $\mathrm{YbB}_{12}$
and a $d$-electron system\cite{Hsieh_TCIinOxides2014} $\mathrm{Sr_3PbO}$.

Recently, effects of interactions on topological crystalline insulators
have been addressed by Isobe and Fu.\cite{Isobe_Fu2015}
Motivated by first-principles calculations\cite{SnTe_2D,Safaei2015}
predicting that thin films of SnTe become two-dimensional topological
insulators (i.e., quantum spin Hall insulators),
they studied stability of gapless edge states with
internal $Z_2$ symmetry that comes from reflection
symmetry about the two-dimensional plane;
they obtained $\mathbb{Z}_4$ classification for thin films.
Furthermore, they have developed a theoretical approach to classify
three-dimensional topological crystalline insulators which
utilizes the stability analysis of the gapless edge states.
This approach led to $\mathbb{Z}_8$ classification for
three-dimensional topological crystalline insulators.
Strictly speaking, however, the symmetry of the models
studied in Ref.~\onlinecite{Isobe_Fu2015} is U(1)$\times Z_2$,
where the U(1) symmetry denotes charge conservation\footnote{
In two dimensions the group structure of the SPT phases with
U(1)$\times Z_2$ symmetry is $\mathbb{Z}\times\mathbb{Z}_4$.
In the noninteracting case each eigensector of $Z_2$ symmetry
(having eigenvalues $+i$ and $-i$ for reflection)
is insulators in class A and
characterized by a (mirror) Chern number;
the topological classification in the noninteracting case
is $\mathbb{Z}\times\mathbb{Z}$.
Under the assumption of the vanishing total Chern number
(removing one $\mathbb{Z}$ from $\mathbb{Z}\times\mathbb{Z}$),
the classification is reduced from $\mathbb{Z}$ to $\mathbb{Z}_4$
by interactions, as shown in Ref.~\onlinecite{Isobe_Fu2015}}.
In Ref.~\onlinecite{Isobe_Fu2015},
the time-reversal symmetry of topological crystalline insulators
is taken into account simply by assuming nonchiral
gapless edge (domain-wall) modes.
While this procedure gives correct classification, in our opinion,
their treatment of time-reversal symmetry is not fully satisfying.

In this paper we study interaction effects on
topological crystalline insulators
protected by time-reversal and reflection symmetries.
That is, we classify fermionic SPT phases under
$Z_2\times[\mbox{U(1)}\rtimes T]$ symmetry in two dimensions
and $R\times[\mbox{U(1)}\rtimes T]$ symmetry in three dimensions,
where $T$ and $R$ denote time-reversal and reflection.
Using the models studied by Isobe and Fu,
we carefully derive transformation laws of fields under symmetry
transformations and examine the stability of gapless edge modes
under symmetry-preserving perturbations and interactions.
We obtain $\mathbb{Z}_4$ classification for two dimensions
and $\mathbb{Z}_8$ classification for three dimensions,
in agreement with Isobe and Fu.

The rest of this paper is organized as follows. 
In Sec.~\ref{sec: two_dimensions}, we consider a two-dimensional model
respecting the reflection symmetry whose reflection plane is parallel
to the two-dimensional plane.
We elucidate collapse of topological classification
from $\mathbb{Z}$ to $\mathbb{Z}_4$ due to interactions. 
In Sec.~\ref{sec: three_dimensions},
by extending the argument of Ref.~\onlinecite{Isobe_Fu2015}
to time-reversal invariant systems,
we show collapse of topological
classification from $\mathbb{Z}$ to $\mathbb{Z}_8$
in three-dimensional topological crystalline insulators.
This means that eight Dirac cones on the surface of a topological crystalline
insulator can be gapped out.
In addition, in Sec.~\ref{sec: topological_order}
we point out that two Dirac cones can be gapped out without symmetry breaking
by attaching a fractionalized quantum spin Hall insulator
with topological order.
In Sec.~\ref{sec. summary} we summarize our results.

\section{two-dimensional topological crystalline insulator}\label{sec: two_dimensions}
Recent first-principles calculations predict that
quantum spin Hall states can be realized in the (111) thin films
of the SnTe class of three-dimensional topological crystalline
insulators.\cite{SnTe_2D,Safaei2015}
The surface Dirac fermions on the top and bottom surfaces of
the (111) thin films are gapped by intersurface coupling,
and turn into a topological state.
Following Ref.~\onlinecite{SnTe_2D}, we take the effective Hamiltonian
for the Dirac fermions on
the top and bottom surfaces of a thin film with an odd number
of layers,
\begin{equation}
H_{2D}=
\int d^2\bm{x}
\bm{\psi}^\dagger(\bm{x})
\left[\left(
i\partial_x\sigma^y-i\partial_y\sigma^x \right)\otimes \tau^z
+m\tau^x \right]
\bm{\psi}(\bm{x}),
\label{H_2D}
\end{equation}
where $\bm{x}=(x,y)$,
$\bm{\psi}^\dagger=
(\psi^\dagger_{1\uparrow},\psi^\dagger_{1\downarrow},
\psi^\dagger_{2\uparrow},\psi^\dagger_{2\downarrow})$,
and the velocity is set equal to unity.
Here $\psi^\dagger_{\alpha\sigma}$ is the creation operator of Dirac fermions
with spin $\sigma=\uparrow,\downarrow$ on the top ($\alpha=1$) or
the bottom ($\alpha=2$) surface.
The Pauli matrices $\sigma^i$ and $\tau^i$ $(i=x,y,z)$
act on the spin and surface indices,
respectively.
The coupling of the top and bottom surface Dirac fermions
gives the mass term $m\tau^x$.

Notice that the Hamiltonian (\ref{H_2D}) respects the time-reversal ($T$)
symmetry and the local $Z_2$ symmetry ($g$) arising from the reflection
symmetry with respect to the two-dimensional plane.
Transformations of $\bm{\psi}$ under the symmetry operations are described by
\begin{equation}
T=-i\sigma^yK,\quad
g=i\sigma^z\tau^x,
\label{T and g on psi}
\end{equation}
where $K$ is the operator for complex conjugation.

The presence of the $Z_2$ symmetry promotes
the topological index characterizing the two-dimensional Hamiltonian
from $\mathbb{Z}_2$
($\mathbb{Z}_2$ index of time-reversal invariant insulators)
to $\mathbb{Z}$ (a mirror Chern number).
To understand this, we first note that
the Chern number is defined for each eigenspace of $g$.
The operators $T$ and $g$
commute with each other,
$[-i\sigma^yK,i\sigma^z\tau^x]=0$,
and the time-reversal symmetry interchanges the eigenspaces of $g$.
For example,
if $|+\rangle$ and $|-\rangle$ are eigenstates of $g$ with the eigenvalues
$+i$ and $-i$, respectively, then $T|\pm \rangle$ are eigenstates
with the eigenvalues $\mp i$, respectively.
Thus, the Hamiltonian is block diagonalized into eigenspaces of $g$
which are related by the time-reversal symmetry
and characterized by the Chern numbers with opposite signs\footnote{
Without the time-reversal symmetry, the two eigensectors
have independent Chern numbers, and the topological classification
is $\mathbb{Z}\times\mathbb{Z}$}.
Having the integer topological number, $N_0$ copies of $H_{2D}$
can have $N_0$ pairs of helical edge modes protected by time-reversal
and reflection symmetries.

To describe a Kramers pair of helical edge modes
of the two-dimensional topological insulator,
we introduce smooth spatial modulation in the mass term in Eq.\ (\ref{H_2D}).
This generates gapless states
moving along a line where the sign of the mass changes.
Let us replace $m\tau^x$ by $m(x)\tau^x$, where $m(x)$ takes a positive
(negative) value for $x>0$ ($x<0$).
This yields helical edge states localized at the kink ($x=0$);
solving the Dirac equation for the zeromode,
\begin{equation}
[i\partial_x\sigma^y\otimes\tau^z + m(x)\tau^x] \,|\mbox{$\pm y$}\rangle
 = 0,
\end{equation}
we obtain a Kramers pair of gapless modes
$|\mbox{$+y$}\rangle$ and $|\mbox{$-y$}\rangle$
propagating to the $+y$ and $-y$ directions, respectively:
\begin{subequations}
\label{helical states}
\begin{equation}
\langle\bm{x} |\mbox{$\pm y$}\rangle =
\exp\!\left[\pm ik_y y -\int^x_0 dx' m(x')\right]\!
|\mbox{$\pm y$}\rangle_0, 
\end{equation}
with 
\begin{equation}
|\mbox{$\pm y$}\rangle_0 =
\left(
\begin{array}{c}
1  \\
i
\end{array}
\right)_{\sigma}
\otimes
\left(
\begin{array}{c}
1  \\
i
\end{array}
\right)_{\tau}
\pm i
\left(
\begin{array}{c}
1  \\
-i
\end{array}
\right)_{\sigma}
\otimes
\left(
\begin{array}{c}
1  \\
-i
\end{array}
\right)_{\tau}
,
\end{equation}
\end{subequations}
where $|\bm{x}\rangle$ is the eigenstate of $\bm{x}$,
and $k_y$ denotes the momentum along the $y$ direction.
These states are transformed as
\begin{equation}
T
\begin{pmatrix}
|\mbox{$+y$}\rangle \\
|\mbox{$-y$}\rangle
\end{pmatrix}
=
\begin{pmatrix}
|\mbox{$-y$}\rangle \\
-|\mbox{$+y$}\rangle
\end{pmatrix},
\quad 
g
\begin{pmatrix}
|\mbox{$+y$}\rangle \\
|\mbox{$-y$}\rangle
\end{pmatrix}
=
\begin{pmatrix}
i\,|\mbox{$+y$}\rangle \\
-i\,|\mbox{$-y$}\rangle
\end{pmatrix}.
\label{T and g on |pm y>}
\end{equation}

In order to examine the stability of the gapless helical edge modes
in the presence of interactions, we bosonize the fermionic gapless
edge modes.\cite{Neupert_CS_2011,Levin_CS_2012,Lu_CS_2011,Hsieh_CS_CPT_2014}
The Lagrandian for the bosonic fields representing the helical
edge modes is given by
\begin{subequations}
\begin{equation}
\mathcal{L}=
\int\!\frac{dx}{4\pi}
\left[
K_{I,J}\partial_t \phi_I(x) \partial_x \phi_J(x)
-\partial_x \phi_I(x) \partial_x \phi_I(x)
\right],
\label{eq: L}
\end{equation}
with $K=\rho^z$, where $\rho^i$ ($i=x,y,z$) are the Pauli matrices,
and summation over the repeated indices $I,J$ is assumed
(this is assumed throughout this paper).
The bosonic fields, $\bm{\phi}=(\phi_1,\phi_2)=(\phi^+,\phi^-)$,
describe the edge modes
propagating to the $+y$ and $-y$ directions, respectively.
More specifically,
the vertex operators $\mbox{$:\!e^{i\phi^+}\!\!:$}$
and $\mbox{$:\!e^{i\phi^-}\!\!:$}$ create
fermions in the $|\mbox{$+y$}\rangle$ and $|\mbox{$-y$}\rangle$ states,
respectively.
The fields $\phi_I$ are defined modulo $2\pi$.
The commutation relations of these fields are given by
\begin{equation}
[\phi_I(x),\phi_J(y)] =
i\pi \rho^z_{I,J}\mathrm{sgn}(x-y) +i\pi\,\mathrm{sgn}(I-J).
\label{eq: [phi_I,phi_J] rho^z}
\end{equation}
\end{subequations}
Here, $\mathrm{sgn}(x)$ equals $+1$, $0$, and $-1$ for $x>0$, $x=0$,
and $x<0$ respectively. 
The second term on the right hand side of 
Eq.~(\ref{eq: [phi_I,phi_J] rho^z}) accounts for the anticommutation 
relation of fermions from different edge modes.
Symmetry transformations of the bosonic fields $\phi_I$ can be
deduced from Eq.\ (\ref{T and g on |pm y>}),
which should be compared with
$\mathcal{G}\,\mbox{$:\!e^{i\phi_{1,2}}\!\!:$}\,\mathcal{G}^{-1}$
($\mathcal{G}=\hat{T}, \hat{g}, \hat{u}_\theta$).
We thus obtain the following transformation rules for the bosonic fields:
\begin{subequations}
\label{eq: trans-Tg_1}
\begin{eqnarray}
\hat{T}\bm{\phi}(x)\hat{T}^{-1}
 &=& -\sigma^x \bm{\phi}(x) +\pi \bm{e}_2, \\
\hat{g}\bm{\phi}(x)\hat{g}^{-1}
 &=& \bm{\phi}(x) +\frac{\pi}{2} (\bm{e}_1-\bm{e}_2), \\
\hat{u}_{\theta} \bm{\phi}(x) \hat{u}^{-1}_{\theta}
 &=& \bm{\phi}(x) +\theta (\bm{e}_1+\bm{e}_2),
\end{eqnarray}
\end{subequations}
where $\bm{e}_I$ is the unit vector whose $I$th entry is one
and the other entries are zero.
The operators $\hat{T}$, $\hat{g}$, and $\hat{u}_{\theta}$ denote
the time-reversal, the local reflection, and the charge U(1) rotation,
respectively.
The minus sign in Eq.~(\ref{eq: trans-Tg_1}a) is due to
the antiunitary nature of $\hat{T}$.
The symmetry group of the system is denoted as
$Z_2\times [\mathrm{U(1)}\rtimes T]$ in
the notation of Ref.~\onlinecite{Lu_CS_2011}.

We will examine whether interactions can gap out
$N_0$ copies of the gapless helical modes defined above
without symmetry breaking for $N_0=1,\cdots,4$.
The Lagrangian for the $2N_0$-component bosonic field
$\bm{\phi}=(\phi_1,\ldots,\phi_{2N_0})^T$, which describes
the $N_0$ copies of helical edge modes,
is given by Eq.~(\ref{eq: L}) with the $K$ matrix
\begin{subequations}
\begin{equation}
K=\rho^z\otimes\1_{N_0},
\end{equation}
where $\1_{N_0}$ is the $N_0\times N_0$ unit matrix.
The $I$th component of the bosonic field $\phi_I$ corresponds to
the edge mode propagating to the $+y$ ($-y$) direction
for odd $I$ (even $I$), respectively.
The commutation relations of the bosonic fields
$\phi_I$ ($I=1,\ldots,2N_0$) are given by
\begin{equation}
[\phi_I(x),\phi_J(y)] =
i\pi\!\left(K^{-1}\right)_{I,J}\mathrm{sgn}(x-y) +i\pi\,\mathrm{sgn}(I-J).
\label{eq: comm}
\end{equation}
\end{subequations}
The transformation rules in Eq.~(\ref{eq: trans-Tg_1})
hold if we replace $\bm{e}_1$ and $\bm{e}_2$ with
the following $2N_0$-dimensional vectors:
$\bm{e}_1=(1,0,1,0,\cdots,1,0)$ and
$\bm{e}_2=(0,1,0,1,\cdots,0,1)$.

The $N_0$ pairs of helical edge modes are gapped out
if there exist potential
\begin{equation}
\mathcal{L}_\mathrm{int}=
\sum_{\alpha}C_{\alpha}\int dx
:\!\cos(\bm{l}_{\alpha}\cdot\bm{\phi}+a_{\alpha})\!:
\label{eq: int_framefork}
\end{equation}
with
$N_0$ linearly independent vectors $\bm{l}_\alpha$ satisfying
the Haldane's null vector condition\cite{haldane_nullvector}
\begin{equation}
\bm{l}^T_{\alpha}K^{-1}\bm{l}_{\beta}=0,
\qquad
(\alpha,\beta=1,\cdots, N_0)
\label{Haldane criteria}
\end{equation}
so that the fields satisfy
$[\bm{l}_\alpha\cdot\bm{\phi}(x),\bm{l}_\beta\cdot\bm{\phi}(y)]=0$
up to $2\pi in$ ($n\in \mathbb{Z}$) for $\alpha,\beta=1,\cdots, N_0$.
The coupling constants $C_\alpha$ and the phases $a_\alpha$
are real numbers.
As indicated by the colons
in Eq.~(\ref{eq: int_framefork}), the vertex operators
are normal-ordered,
\begin{equation}
:\!e^{i\bm{l}\cdot\bm{\phi}}\!: \,
=
e^{i\frac{\pi}{2}\sum_{I<J}l_Il_J}
:\!e^{il_1\phi_1}\!: \,
:\!e^{il_2\phi_2}\!: \cdots
:\!e^{il_{2N_0}\phi_{2N_0}}\!\!: \, ,
\label{eq: normal_product}
\end{equation}
where the phase factor $\exp(\frac{i\pi}{2}\sum_{I<J}l_Il_J)$ is
in accordance with the second term in the commutator in Eq.\ (\ref{eq: comm}).
It makes $\mbox{$:\!e^{-i\bm{l}\cdot\bm{\phi}}\!:$}$ the Hermitian conjugate
of $\mbox{$:\!e^{i\bm{l}\cdot\bm{\phi}}\!:$}$.

There are cases when the symmetry is broken spontaneously,
even when $\mathcal{L}_\mathrm{int}$ in 
Eq.~(\ref{eq: int_framefork}) is invariant under the symmetry
transformations.
The occurrence of spontaneous symmetry breaking can be
judged by finding elementary bosonic variables as follows.
The $N_0$ linearly independent vectors
$\{\bm{l}_1,\cdots,\bm{l}_{N_0}\}$
form a $N_0$-dimensional lattice 
as $\bm{R}=j_1\bm{l}_1+j_2\bm{l}_2+\cdots+j_{N_0}\bm{l}_{N_0}$,
where 
$j_\alpha$ are arbitrary integers ($\alpha=1,\ldots,N_0$).
From the integer vectors in the lattice,
we find a set of linearly independent vectors
$\{ \tilde{\bm{l}}_1,\tilde{\bm{l}}_2,\cdots,\tilde{\bm{l}}_{N_0} \}$,
from which we define a primitive lattice vectors
$\{\bm{v}_1,\cdots, \bm{v}_{N_0}\}$,
\begin{equation}
\bm{v}_{\alpha}=
\frac{1}{\mathrm{gcd}(\tilde{l}_{\alpha,1},\cdots, \tilde{l}_{\alpha,2N_0})}
\tilde{\bm{l}}_{\alpha} ,
\end{equation}
where
gcd denotes the greatest common divisor of the integers
in the parentheses.
The vectors $\{\bm{v}_1,\cdots,\bm{v}_{N_0}\}$
form a primitive cell of the smallest volume
(see examples below).
Finally, elementary bosonic variables\cite{Lu_CS_2011}
are given by
$\bm{v}_\alpha \cdot \bm{\phi}$ ($\alpha=1,\cdots,N_0$).
The set of elementary bosonic variables
$\{ \bm{v}_1 \cdot \bm{\phi},\cdots, \bm{v}_{N_0} \cdot \bm{\phi} \}$
are pinned to constant values, when the fields
$\{ \bm{l}_1 \cdot \bm{\phi},\cdots, \bm{l}_{N_0} \cdot \bm{\phi} \}$
are pinned by the potentail in $\mathcal{L}_\mathrm{int}$.
We see that
the gapless edge modes can be gapped out without symmetry breaking
if and only if the set
$\{ \bm{v}_1\cdot \bm{\phi},\cdots, \bm{v}_{N_0}\cdot\bm{\phi} \}$
is invariant,
\begin{equation}
\mathcal{G}
\{ \bm{v}_1\cdot \bm{\phi},\cdots, \bm{v}_{N_0}\cdot\bm{\phi} \}
\mathcal{G}^{-1}
=
\{ \bm{v}_1\cdot \bm{\phi},\cdots, \bm{v}_{N_0}\cdot\bm{\phi} \},
\end{equation}
modulo $2\pi$
under the symmetry transformations in Eq.\ (\ref{eq: trans-Tg_1}),
where $\mathcal{G}=\hat{T}, \hat{g}, \hat{u}_{\theta}$.

Here, two comments on the symmetry transformations
of the cosine terms are in order.
First, the vector $\bm{l}_0=K^{-1}(1,1,\cdots,1)^T=(1,-1,\cdots,1,-1)$
is in the lattice
generated by the elementary vectors $\{ \bm{v}_1,\cdots, \bm{v}_{N_0}\}$,
when the cosine terms with $\{ \bm{l}_1,\cdots, \bm{l}_{N_0} \}$
in $\mathcal{L}_\mathrm{int}$ respect the symmetry
and gap out all the edge modes. 
In other words, the field $\bm{l}_0\cdot\bm{\phi}$ is pinned
when all the edge modes are gapped out.
This is understood by noting that
$\mbox{$:\!\cos(2\bm{l}_0\cdot\bm{\phi})\!:$}$
is invariant under the symmetry transformations.
Second, it is important to take into account additional phase shifts
coming from the Klein factors [the second term in
Eq.\ (\ref{eq: int_framefork}a)] when we examine the invariance
of the cosine terms in $\mathcal{L}_\mathrm{int}$,
in particular, under the time-reversal
transformation, which changes the direction of propagation of edge modes.

We show below that four copies of the helical edge modes (\ref{eq: L})
can be gapped out without symmetry breaking. 
We begin with the case of $N_0=1$.
A gapping potential respecting the symmetry is given by
\begin{subequations}
\label{eq: int_single}
\begin{equation}
\mathcal{L}_{int}=
C\int dx :\!\cos\left[2\bm{v}\cdot \bm{\phi}(x) \right]\!:
\end{equation}
with 
\begin{equation}
\bm{v}=\left(1,-1\right)^T.
\end{equation}
\end{subequations}
This potential is invariant under the operations of $\hat{T}$, $\hat{g}$,
and $\hat{u}_{\theta}$. 
Indeed, the invariance under $\hat{g}$ and $\hat{u}_{\theta}$ can be seen
as 
\begin{subequations}
\label{eq: trans_2ebv}
\begin{eqnarray}
\hat{g} \left(2\bm{v}\cdot\bm{\phi}\right) \hat{g}^{-1}
&=& 2\bm{v}\cdot\bm{\phi}+2\pi, \\
\hat{u}_{\theta} \left(2\bm{v}\cdot\bm{\phi}\right)\hat{u}^{-1}_{\theta}
&=& 2\bm{v}\cdot\bm{\phi}.
\end{eqnarray}
\end{subequations}
To verify the invariance under $\hat{T}$, we note
from Eqs.\ (\ref{eq: trans-Tg_1}a) and (\ref{eq: normal_product}) that
\begin{subequations}
\begin{eqnarray}
\hat{T}:\!e^{2i\bm{v}\cdot\bm{\phi}}\!:\hat{T}^{-1}
&=&
\hat{T}:\!e^{2i\phi_1}\!:\,:\!e^{-2i\phi_2}\!:\hat{T}^{-1}
\nonumber\\
&=&
:\!e^{2i\phi_2}\!:\,:\!e^{-2i\phi_1}\!:
\nonumber\\
&=&
:\!e^{-2i\bm{v}\cdot\bm{\phi}}\!:
\end{eqnarray}
and vice versa, and hence
\begin{equation}
\hat{T}:\!\cos(2\bm{v}\cdot\bm{\phi})\!:\hat{T}^{-1}
=:\!\cos(2\bm{v}\cdot\bm{\phi})\!:.
\end{equation}
\end{subequations}
It turns out, however, that the ground state breaks the symmetry
as the elementary bosonic variable $\bm{v}\cdot\bm{\phi}$ is transformed as
\begin{equation}
\hat{g}\, \bm{v}\cdot \bm{\phi}\, \hat{g}^{-1}
= \bm{v}\cdot \bm{\phi} +\pi.
\label{eq: g_to_ebv}
\end{equation}
The vector $\bm{v}$ in Eq.~(\ref{eq: int_single}b) is
the elementary bosonic variable respecting the U(1) symmetry.
Hence, it is impossible to gap out edge modes by any symmetry-preserving
potential without spontaneous symmetry breaking
when $N_0=1$.

For $N_0=2$, an example of pinning potentials allowed by the symmetry
is given by
\begin{subequations}
\begin{equation}
\mathcal{L}_\mathrm{int}
= C\sum_{\alpha=1,2} \int dx
 :\!\cos[2\bm{v}_{\alpha}\cdot \bm{\phi}(x)]\!:
\end{equation}
with
\begin{equation}
\bm{v}_1=\left(1,0\, |\, 0,-1\right), 
\quad
\bm{v}_2=\left(0,1\, |-1,0\right),
\end{equation}
\end{subequations}
where $C$ is an arbitrary real number,
and the vertical lines in $\bm{v}_{1,2}$ are inserted between the copies of
helical edge modes. 
The invariance under $\hat{g}$ and $\hat{u}_{\theta}$
can be checked directly as in the $N_0=1$ case.
The time-reversal transformation of
$\bm{\phi}=(\phi_1,\phi_2,\phi_3,\phi_4)^T$,
\begin{equation}
\hat{T}\bm{\phi} \hat{T}^{-1}=
-\sigma^x\oplus\sigma^x \bm{\phi} +\pi (\bm{e}_2+\bm{e}_4),
\end{equation}
yields
\begin{subequations}
\begin{eqnarray}
\hat{T} :\!\cos(2\bm{v}_1\cdot\bm{\phi})\!:\hat{T}^{-1}
&=& :\!\cos(2\bm{v}_2\cdot\bm{\phi})\!: \, , \\
\hat{T} :\!\cos(2\bm{v}_2\cdot\bm{\phi})\!:\hat{T}^{-1}
&=& :\!\cos(2\bm{v}_1\cdot\bm{\phi})\!: \, ,
\end{eqnarray}
\end{subequations}
where we have used Eq.~(\ref{eq: normal_product}). 
However, the spontaneous symmetry breaking occurs
once the elementary bosonic variables
$\{ \bm{v}_1\cdot\bm{\phi},\bm{v}_2\cdot\bm{\phi} \}$
are pinned by $\mathcal{L}_\mathrm{int}$,
as these variables are not invariant under $\hat{g}$
[in a similar way to Eq.~(\ref{eq: g_to_ebv})]. 
We note that this observation holds for arbitrary $\bm{l}$'s
of the pinning potentials respecting the U(1) symmetry.
For example, if the cosine terms with
\begin{subequations}
\begin{eqnarray}
\bm{l}_1&=& (1,-1 \,|\, 1,-1)^T, \\
\bm{l}_2&=& (1,1 \,| -1,-1)^T,
\end{eqnarray}
\end{subequations}
are used to gap out the edge modes,
then the elementary bosonic fields
$\{\bm{v}_1\cdot\bm{\phi},\bm{v}_2\cdot\bm{\phi}\}$
are also pinned (because
$\bm{l}_1+\bm{l}_2=2\bm{v}_1$ and $\bm{l}_1-\bm{l}_2=-2\bm{v}_2$),
which gives rise to spontaneous symmetry breaking.
Thus, we cannot gap out edge modes without symmetry breaking. 

Let us move on to the case of $N_0=3$. 
When all the helical edge modes are gapped out, the field 
\begin{eqnarray}
\bm{l}_0\cdot\bm{\phi}&=& \phi_1-\phi_2+\phi_3-\phi_4+\phi_5-\phi_6,
\end{eqnarray}
necessarily takes a classical value
(i.e., $\langle \bm{l}_0\cdot\bm{\phi} \rangle=\mathrm{const}.$).
This implies that the ground state spontaneously breaks the symmetry,
as $\bm{l}_0\cdot\bm{\phi}$ is not invariant under $\hat{g}$,
\begin{equation}
\hat{g}\,\bm{l}_0\cdot\bm{\phi}\,\hat{g}^{-1}=\bm{l}_0\cdot\bm{\phi}+3\pi.
\end{equation}

For $N_0=4$, we can gap out all the edge modes while respecting the symmetry
by taking the following cosine terms:
\begin{subequations}
\label{cosine potentail for N_0=4}
\begin{equation}
\mathcal{L}^{(4)}_\mathrm{int}=
C\sum_{\alpha=1}^4\int dx
:\!\cos\left[\bm{v}_{\alpha}\cdot\bm{\phi}(x)\right]\!:
\end{equation}
with
\begin{eqnarray}
\bm{v}_1&=& \left(1,0 \,|\, 1,0 \,|\, 0,-1 \,|\, 0,-1\right)^T,  \\
\bm{v}_2&=& \left(0,1\, |\, 0,1 \,| -1,0 \,| -1,0\right)^T,  \\
\bm{v}_3&=& \left(1,-1 \,| -1,1 \,|\, 0,0 \,|\, 0,0\right)^T,  \\
\bm{v}_4&=& \left(0,0 \,|\, 0,0 \,|\, 1,-1 \,| -1,1\right)^T.
\end{eqnarray}
\end{subequations}
The time-reversal invariance can be seen as follows:
\begin{subequations}
\begin{eqnarray}
\hat{T}:\!\cos(\bm{v}_1\cdot\bm{\phi})\!:\hat{T}^{-1}
&=&:\!\cos(\bm{v}_2\cdot\bm{\phi})\!: \,, \\
\hat{T}:\!\cos(\bm{v}_2\cdot\bm{\phi})\!:\hat{T}^{-1}
&=&:\!\cos(\bm{v}_1\cdot\bm{\phi})\!: \,, \\
\hat{T}:\!\cos(\bm{v}_3\cdot\bm{\phi})\!:\hat{T}^{-1}
&=&:\!\cos(\bm{v}_3\cdot\bm{\phi})\!: \,, \\
\hat{T}:\!\cos(\bm{v}_4\cdot\bm{\phi})\!:\hat{T}^{-1}
&=&:\!\cos(\bm{v}_4\cdot\bm{\phi}):\!.
\end{eqnarray}
\end{subequations}
Invariance under the transformations of $\hat{g}$ and $\hat{u}_{\theta}$
can be seen by straightforward calculations;
\begin{subequations}
\label{eq: trans_gu_2}
\begin{eqnarray}
\hat{g}\,\bm{v}_{\alpha}\cdot \bm{\phi} \,\hat{g}^{-1}
&=& \bm{v}_{\alpha}\cdot \bm{\phi}  \quad (\mathrm{mod} \; 2\pi), \\
\hat{u}_{\theta}\bm{v}_{\alpha}\cdot \bm{\phi} \,\hat{u}^{-1}_{\theta}
&=& \bm{v}_{\alpha}\cdot \bm{\phi},
\end{eqnarray}
\end{subequations}
for $\alpha=1,\cdots,4$.
When elementary bosonic variables
$\bm{v}_\alpha\cdot \bm{\phi}$ $(\alpha=1,\cdots,4)$ are pinned
at some constant values such that
$\langle\bm{v}_1\cdot\bm{\phi}\rangle=\langle\bm{v}_2\cdot\bm{\phi}\rangle$,
the edge modes are completely gapped out without symmetry breaking.
Hence, we conclude that two-dimensional SPT phases form
a $\mathbb{Z}_4$ group,
which is reduced from $\mathbb{Z}$ of the noninteracting fermions.

\section{Three-dimensional topological insulator} \label{sec: three_dimensions}
In this section, we discuss three-dimensional topological insulators with
the time-reversal and reflection symmetry.
In Ref.~\onlinecite{Isobe_Fu2015}, Isobe and Fu
studied interaction effects on topological crystalline insulators
characterized by mirror Chern numbers.
They introduced a spatially varying mass term to the surface Dirac
Hamiltonian and examined the stability of one-dimensional
gapless modes propagating along domain walls where the sign of
the Dirac mass changes.
We note that the topological crystalline insulating phase in
$\mathrm{SnTe}$ is an SPT phase protected by both reflection and
time-reversal symmetries.
In Ref.~\onlinecite{Isobe_Fu2015}, however, the time-reversal symmetry is
not explicitly considered and is used only in the assumption
of the nonchiral structure of boundary (domain wall) states.%
~\footnote{
In the classification sheme based on K theory,
one finds that the relevant classifying space depends on
whether one imposes additional reflection symmetry on class A
(without time-reversal
symmetry) or class AII (with time-reversal symmetry).
In the former case the relevant classifying space is $C_0$
while it is $R_0$ in the latter case\cite{Morimoto_classification2013}}
Here, we improve on their approach by keeping time-reversal symmetry
intact at every step of calculations and show that topological classification
collapse from $\mathbb{Z}$ to $\mathbb{Z}_8$.

Let us start with discussion on the topological invariant
in the noninteracting case.
As is well known, topological phases of class AII in three dimensions
form a $\mathbb{Z}_2$ group. 
In the presence of additional reflection symmetry,
the group structure of the topological phases is promoted to
$\mathbb{Z}$.\cite{Hsieh_TCH_SnTe_2012}
When the reflection plane is the $yz$-plane,
the reflection operator is given by
\begin{equation}
R=i\sigma^xP,
\end{equation}
where $\sigma^x$ acts on the electron spin,
and $P$ changes the sign of $x$ coordinate, $P:(x,y,z)\to(-x,y,z)$. 
The relevant symmetry group is denoted as
$R\times [\mathrm{U(1)} \rtimes T]$ in the notation of
Ref.~\onlinecite{Lu_CS_2011}.
The topological invariant for insulating phases under this symmetry group
is the mirror Chern number $n_{M}=(n^+-n^{-})/2\in\mathbb{Z}$,
where $n^{\pm}$ are the Chern numbers defined in the eigenspaces
of $R$ on the two-dimensional fixed plane under the reflection
in the momentum space.
The time-reversal symmetry $T=-i\sigma^yK$ is not closed
in each eigenspace of $R$ on the fixed plane.

Following Ref.~\onlinecite{Isobe_Fu2015},
we classify the correlated topological crystalline insulators
by introducing a spatially varying Dirac mass term to the surface Hamiltonian.
We will show below that the gapless modes from eight copies of
Dirac cones can be gapped out
without symmetry breaking.
The classification of three-dimensional topological crystalline insulators
is reduced from $\mathbb{Z}$ to $\mathbb{Z}_8$ by interactions.

The Hamiltonian of a single Dirac cone
on the surface of a topological crystalline insulator is given by
\begin{equation}
H_\mathrm{surf}=
\int d^2\bm{x}
\bm{\psi}^\dagger(\bm{x})
(i\partial_x \sigma^y -i\partial_y \sigma^x )
\bm{\psi}(\bm{x}),
\label{eq: single_dirac_three_dim}
\end{equation}
where $\bm{\psi}^\dagger=(\psi^\dagger_\uparrow,\psi^\dagger_\downarrow )$.
Without reflection symmetry,
two copies of the surface Dirac cones are gapped out
without breaking the time-reversal symmetry
by introducing a mass term in the Hamiltonian
\begin{equation}
H^{(2)}_\mathrm{surf}= \!
\int\!d^2\bm{x}\, \bm{\psi}^\dagger(\bm{x}) 
[(i\partial_x\sigma^y -i\partial_y \sigma^x )\otimes\tau_0
+ m_0 \sigma^z\otimes\tau^y ] \bm{\psi}(\bm{x}).
\label{eq: Dirac_mass_RtTrtU}
\end{equation}
Here $\tau$ acts on the pseudospin space in which the eigenvalues
$\pm1$ of $\tau^z$ distingsuish the two Dirac cones,
and $\bm{\psi}^\dagger$ is a four component vector,
$\bm{\psi}^\dagger=(\psi^\dagger_{1\uparrow},\psi^\dagger_{1\downarrow},
\psi^\dagger_{2\uparrow},\psi^\dagger_{2\downarrow})$,
where $\psi^\dagger_{\alpha,\sigma}$ is the creation operator
of a Dirac fermion
with spin $\sigma=\uparrow,\downarrow$ and pesudospin $\alpha=1,2$.
The instability of two Dirac cones is consistent with the
$\mathbb{Z}_2$ classification of symmetry class AII.
However, when the reflection symmetry is imposed,
the mass term is not allowed by the symmetry
($R\, m_0 \sigma^z\otimes\tau^y R^{-1} = -m_0 \sigma^z\otimes\tau^y$),
and the two Dirac cones remain gapless;
the $\mathbb{Z}$ classification follows.

\begin{figure}[t]
\begin{center}
\includegraphics[width=\hsize]{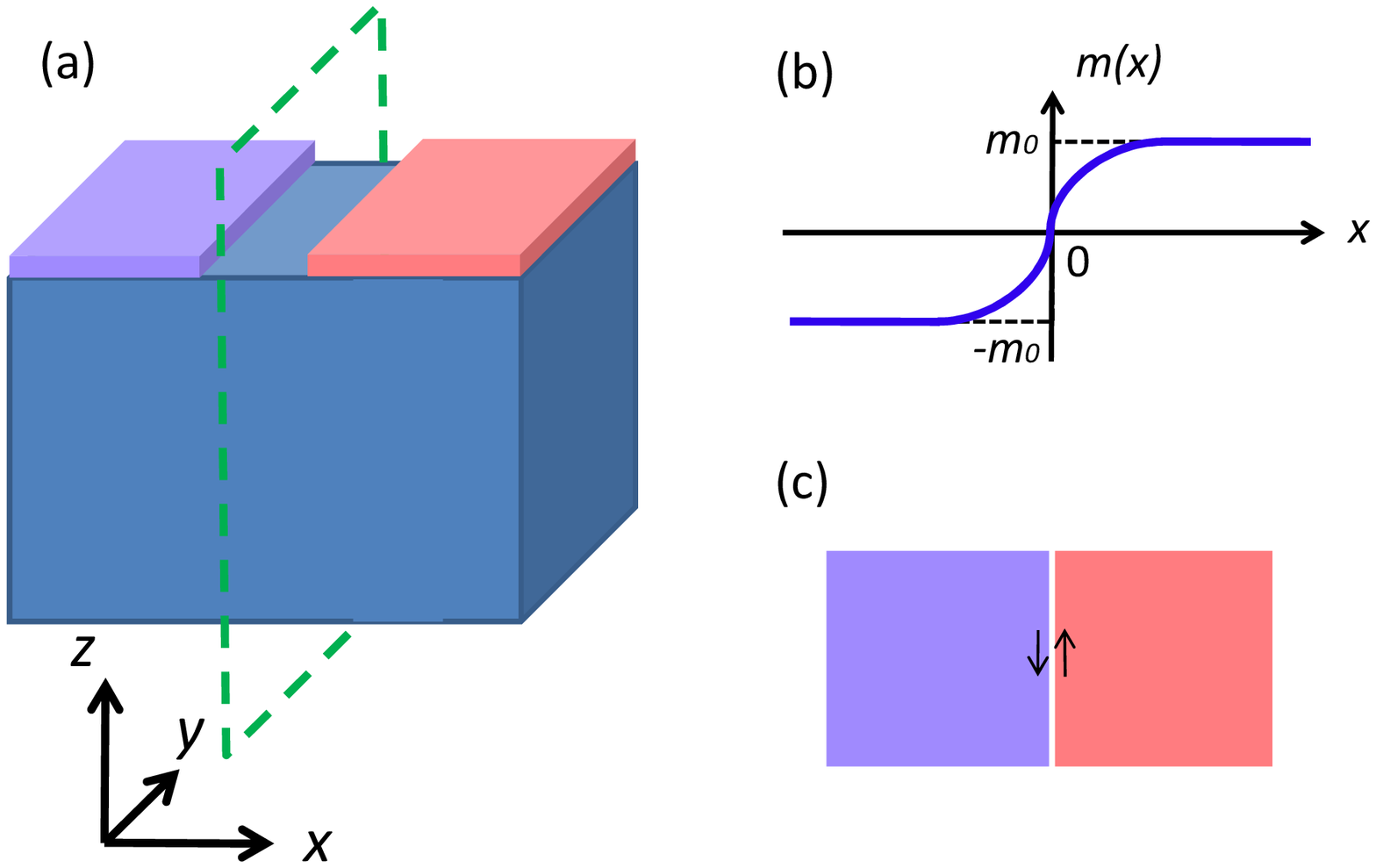}
\end{center}
\caption{
(a) Sketch of a spatially varying Dirac mass introduced to
surface Dirac fermions
of a three-dimensional topological crystalline insulator.
The green dashed square denotes the reflection plane.
We introduce a positive (negative) mass in the region
where the red (violet) thin rectangle is attached.
(b) Spatial dependence of the mass in Eq.~(\ref{eq: Dirac_mass_RtTrtU1}).
(c) Top view of the panel (a).
Helical edge modes are localized at the kink $x=0$.
}
\label{fig: domain}
\end{figure}

One possible way to respect both time-reversal and reflection symmetries
while introducing a mass to Dirac fermions is to modulate the sign of
the mass in space.
This gives rise to one-dimensional gapless modes at the domain wall
where the mass changes its sign.
To see this, let us consider the following simple model:
\begin{eqnarray}
\widetilde{H}^{(2)}_\mathrm{surf}
&=& \int d^2\bm{x}\, \bm{\psi}^\dagger(\bm{x}) 
[(i\partial_x\sigma^y -i\partial_y \sigma^x)\otimes\tau_0 \nonumber \\
&& \qquad\qquad\quad
+ m(x) \sigma^z\otimes\tau^y \,]\bm{\psi}(\bm{x}),
\label{eq: Dirac_mass_RtTrtU1}
\end{eqnarray}
where $m(x)$ takes a positive (negative) value for $x>0$ ($x<0$), respectively.
A sketch of the model is shown in Fig.~\ref{fig: domain}.
When the mass changes its sign as drawn in Fig.~\ref{fig: domain}(b),
a Kramers pair of gapless states propagating along the $y$ direction are
present at the domain wall at $x=0$ [see Fig.~\ref{fig: domain}(c)].
Their wave functions are written as
\begin{subequations}
\label{domain wall states}
\begin{equation}
\langle\bm{x}|\,\mbox{$\pm y$}\rangle =
\exp\!\left[\pm ik_yy - \int^x_0 dx' m(x')\right]
|\,\mbox{$\pm y$}\rangle_0,
\end{equation}
with 
\begin{eqnarray}
|\,\mbox{$+y$}\rangle_0
&=&
i
\begin{pmatrix}
1 \\ 1
\end{pmatrix}_\sigma
\otimes
\begin{pmatrix}
1 \\ i
\end{pmatrix}_\tau,
\\ 
|\,\mbox{$-y$}\rangle_0
&=&
i
\begin{pmatrix}
1 \\ -1
\end{pmatrix}_\sigma
\otimes
\begin{pmatrix}
1 \\ -i
\end{pmatrix}_\tau,
\end{eqnarray}
\end{subequations}
where $|\,\mbox{$+y$}\rangle$ and $|\,\mbox{$-y$}\rangle$
propagate to the $+y$ and $-y$ directions, respectively.
Again, $|\bm{x}\rangle$ denotes the eigenstate of $\bm{x}$.

Under the operations of the time-reversal and reflection symmetry,
these states are transformed as follows:
\begin{subequations}
\begin{equation}
T\,|\,\mbox{$+y$}\rangle = |\,\mbox{$-y$}\rangle,
\quad
T\,|\,\mbox{$-y$}\rangle = - |\,\mbox{$+y$}\rangle,
\end{equation}
and 
\begin{equation}
R\,|\,\mbox{$+y$}\rangle = i |\,\mbox{$+y$}\rangle,
\quad
R\,|\,\mbox{$-y$}\rangle = -i |\,\mbox{$-y$}\rangle.
\end{equation}
\end{subequations}
Hence, introducing the bosonic fields $\phi^{\pm}$ for the gapless
modes propagating along the $\pm y$ directions,
we obtain the transformation laws of $\phi^\pm$ which are identical
to Eqs.\ (\ref{eq: trans-Tg_1}) with $\bm{\phi}=(\phi^{+},\phi^{-})^T$.
Thus, proceeding in the same way as in the previous section,
we see that four copies of gapless modes in
Eq.~(\ref{domain wall states})
can be gapped out
without symmetry breaking by introducing cosine potentials
similar to Eq.\ (\ref{cosine potentail for N_0=4})
that pin the gapless bosonic fields.
Since the helical edge modes in Eq.\ (\ref{domain wall states})
are obtained by introducing a domain wall in the Dirac mass
that couples two Dirac cones,
we conclude that the SPT phases of three-dimensional topological
crystalline insulators protected by time-reversal and reflection
form a $\mathbb{Z}_8$ group, which is reduced from the $\mathbb{Z}$ group
of the noninteracting fermions.

\section{Instability from coupling to topological ordered phases}\label{sec: topological_order}
So far, we have discussed the possibility of gapping out
boundary modes of two- and three-dimensonal topological
crystalline insulators
without symmetry breaking by interactions among themselves.
In this section, we consider a different situation in which
gapless edge modes of a two-dimensional topological crystalline insulator
are interacting with edge modes of a fractionalized
quantum spin Hall insulator (QSHI$^*$).\cite{Lu_Lee2014}
In this case we find that even a single Kramers pair of edge modes
in Eq.~(\ref{helical states}) can be gapped out,
without spontaneous breaking of time-reversal and reflection symmetry.
This result implies that two copies of Dirac cones on the surface
of a three-dimensional topological crystalline insulator can also be
gapped out without symmetry breaking by the coupling to
edge modes of a QSHI$^*$.
In fact, gapped edge states between a QSHI and a QSHI$^*$ are discussed
by Lu and Lee in Ref.~\onlinecite{Lu_Lee2014}.
Here we show that the same symmetric gapped states are obtained
in the presence of additional reflection symmetry.

The QSHI$^{*}$ is obtained by gauging the symmetry of the fermion number
parity from a QSHI;
see Appendix~\ref{sec: app}.
The Lagrangian of edge modes of a QSHI$^*$ is given by
\begin{subequations}
\label{eq: QSHIg}
\begin{equation}
\mathcal{L}=
\int\!\frac{dx}{4\pi}
\left(-2\rho^x_{I,J} \partial_t \phi_I\partial_x \phi_J
 -V_{I,J}\partial_x \phi_I\partial_x \phi_J \right),
\end{equation}
with 
\begin{equation}
\bm{\phi}=(\phi^s,\phi^c)^T,
\end{equation}
\end{subequations}
where $V$ is a symmetric and positive definite matrix. 
The bosonic fields $\phi_c$ and $\phi_s$ describe chargeon and spinon,
respectively. 
These bosonic fields are transformed as follows: 
\begin{subequations}
\label{eq:Tg_QSHIg}
\begin{eqnarray}
\hat{T}
\left(
\begin{array}{c}
\phi^s  \\
\phi^c 
\end{array}
\right)
\hat{T}^{-1}
&=&
\left(
\begin{array}{c}
\phi^s  \\
-\phi^c 
\end{array}
\right)
+
\frac{\pi}{2}
\left(
\begin{array}{c}
-1 \\
1
\end{array}
\right),
\\
\hat{g}
\left(
\begin{array}{c}
\phi^s  \\
\phi^c 
\end{array}
\right)
\hat{g}^{-1}
&=&
\left(
\begin{array}{c}
\phi^s  \\
\phi^c 
\end{array}
\right)
+
\frac{\pi}{2}
\left(
\begin{array}{c}
1  \\
0
\end{array}
\right),
\\
\hat{u}_{\theta}
\left(
\begin{array}{c}
\phi^s  \\
\phi^c 
\end{array}
\right)
\hat{u}_{\theta}^{-1}
&=&
\left(
\begin{array}{c}
\phi^s  \\
\phi^c 
\end{array}
\right)
+
\theta
\left(
\begin{array}{c}
0  \\
1
\end{array}
\right).
\end{eqnarray}
\end{subequations}
Thus, the total Lagrangian of the helical edge modes
at the interface of a two-dimensional topological insulator
(with a mirror Chern number) and a QSHI$^*$ is found
from Eqs.\ (\ref{eq: L}) and (\ref{eq: QSHIg})
to be given by
\begin{subequations}
\label{eq: SPT_QSHI*}
\begin{equation}
\mathcal{L}=\int\! \frac{dx}{4\pi}
\left[K_{I,J} \partial_t \phi_I\partial_x \phi_J
 -V'_{I,J} \partial_x \phi_I\partial_x \phi_J \right], 
\end{equation}
with 
\begin{equation}
K=(-2\rho^x)\oplus\rho^z, \quad V'=V\oplus (v\1_{2}), 
\end{equation}
and 
\begin{equation}
\bm{\phi}= \left(\phi^s,\phi^c,\phi^+,\phi^-\right)^T, 
\end{equation}
\end{subequations}
where $\1_2$ is the two-dimensional identity matrix, and $v$
is the velocity.
We seek cosine potentials of the bosonic fields
$:\!\cos(\bm{l}\cdot \bm{\phi}+\alpha_{\bm{l}})\!:$
that can gap out the edge modes.
The cosine potentials should satisfy
the Haldane's null vector condition in Eq.\ (\ref{Haldane criteria}).
In addition, their $l$-vectors
should be of the form $K\bm{l}'$ with $\bm{l}'\in \mathbb{Z}^4$,
since the cosine potentials are composed of bosonic or fermionic
vertex operators.
The following potential terms satisfy these conditions:
\begin{eqnarray}
\mathcal{L}_{int} &=&
C\int \!dx \left[:\!\cos(2\phi^c-\phi^+-\phi^-)\!:
 \right. \nonumber\\
&&\qquad\qquad \left.
-:\!\cos(2\phi^s -\phi^++\phi^-)\!:\right],
\label{eq: QSHIg_cos}
\end{eqnarray}
where $C$ is a negative constant.
These terms respect all the symmetry we impose: time-reversal, reflection,
and U(1) charge.
Invariance under $\hat{g}$ or $\hat{u}_{\theta}$ operation can be seen
by straightforward calculations, and the time-reversal invariance
can be seen as follows:
\begin{subequations}
\label{eq: QSH* potential_trans}
\begin{eqnarray}
\hat{T}:\!e^{i(2\phi^c-\phi^+-\phi^-)}\!\!:\hat{T}^{-1}
&=&:\! e^{i(2\phi^c-\phi^+-\phi^-)}\!: \, , \\
\hat{T}:\!e^{i(2\phi^s-\phi^++\phi^-)}\!\!:\hat{T}^{-1}
&=&:\!e^{-i(2\phi^s-\phi^++\phi^-)}\!:.
\end{eqnarray}
\end{subequations}
In contrast to the $N_0=2$ case discussed in Sec.~\ref{sec: two_dimensions},
spontaneous symmetry breaking does not occur in the present case.
The elementary vectors associated with the $\bm{l}$ vectors of
the cosine potentials in Eq.\ (\ref{eq: QSHIg_cos})
are given by
\begin{subequations}
\label{eq: QSH*_v}
\begin{eqnarray}
\bm{v}_1&=&(-1,-1,1,0)^T , \\
\bm{v}_2&=&(1,-1,0,1)^T . 
\end{eqnarray}
\end{subequations}
Applying the operator $\hat{T}$ transforms the vertex operators as
\begin{subequations}
\label{eq: QSH*_v T}
\begin{eqnarray}
\hat{T}:\!e^{i\bm{v}_1\cdot \bm{\phi} }\!:\hat{T}^{-1}&=&
-:\!e^{i\bm{v}_2\cdot \bm{\phi} }\!: \, , \\
\hat{T}:\!e^{i\bm{v}_2\cdot \bm{\phi} }\!:\hat{T}^{-1}&=&
-:\!e^{i\bm{v}_1\cdot \bm{\phi} }\!:,
\end{eqnarray}
\end{subequations}
which implies that time-reversal symmetry is preserved
as long as the pinned fields satisfy the relation
\begin{equation}
\langle \bm{v}_1\cdot\bm{\phi}\rangle
-\langle\bm{v}_2\cdot\bm{\phi}\rangle=\pi.
\end{equation}
This is indeed the case with the cosine potentials with $C<0$
in Eq.~(\ref{eq: QSHIg_cos}).
The invariance of the elementary bosonic variables $\bm{v}_i\cdot\bm{\phi}$
under $\hat{g}$ or $\hat{u}_{\theta}$ can also be verified easily. 
Thus, we conclude that the helical edge modes between a QSHI
and a QSHI$^*$ are gapped without symmetry breaking
even in the presence of reflection symmetry.

In Sec.~\ref{sec: three_dimensions} we have considered
a Kramers pair of gapless mode induced at a domain wall
of a spatially varying Dirac mass that couples two Dirac
cones on the surface of a three-dimensional topological
crystalline insulator.
As a corollary of the above result, we conclude that
a gapless helical domain-wall mode is gapped out without symmetry breaking
when coupled to a helical edge mode of a QSHI$^*$.

\section{Conclusion}\label{sec. summary}
Motivated by theoretical proposals of topological crystalline insulators
in correlated electron systems, we have studied fermionc SPT phases
which respect the time-reversal and reflection symmetry
by employing the Chern-Simons approach.
For two-dimensional systems respecting the reflection symmetry
whose reflection plane is parallel to the two-dimensional plane,
our analysis elucidates that the SPT phases form $\mathbb{Z}_4$ group,
while topological classification in the noninteracting case is $\mathbb{Z}$.
Furthermore, we have addressed classification of three-dimensional
topological crystalline insulators by extending the argument proposed
in Ref.~\onlinecite{Isobe_Fu2015} with keeping time-reversal symmetry intact.
Our analysis revealed that eight Dirac cones on the surface
are completely gapped out without symmetry breaking
by two-particle backscattering of gapless modes on the domain wall
of varying Dirac mass.
This leads to collapse of topological classification
from $\mathbb{Z}$ to $\mathbb{Z}_8$. 
Finally, we have pointed out the instability of gapless boundary
modes through coupling to topological ordered phases.

\begin{acknowledgments}
This work was in part supported by Grant-in-Aid from the Japan Society for
Promotion of Science (Grant No.~15K05141)
and by the RIKEN iTHES Project.
\end{acknowledgments}

\appendix*
\section{GAUGING FERMION PARITY SYMMETRY TO FERMIONIC SPT PHASES}\label{sec: app}
In this appendix we review the fractionalized quantum spin Hall insulator,
QSHI$^{*}$, which is obtained by gauging symmetry of the fermion number
parity $P_f$ of a QSHI.\cite{LuVishwanath_gauge2013,Lu_Lee2014} 
The effective action of the QSHI is written as
\begin{equation}
\mathcal{L}_\mathrm{CS}=
\int d^2\bm{x}\left(
\frac{\epsilon^{\mu\nu\rho}}{4\pi}K_{I,J}
a^I_\mu\partial_\nu a^J_\rho -j^\mu l_I a^I_\mu
\right),
\label{eq: CS0_bulk}
\end{equation}
with $K=-\rho^z$.
Here, $\epsilon^{\mu\nu\rho}$ is the anti-symmetric tensor
($\epsilon^{012}=1$, $\mu,\nu,\rho=0,1,2$),
$\bm{\partial}:=(\partial_t,\partial_{x},\partial_{y})$,
$\bm{x}:=(x,y)$.
The internal Chern-Simons gauge fields $a_{I\mu}$ ($I=1,2$) describe
the bulk low-energy excitations in the QSHI.
Summation over repeating indices is assumed. 
The integer vector $\bm{l}\in\mathbb{Z}^2$ characterizes quasiparticles,
whose equation of motion is given by 
\begin{equation}
l_Ij^{\mu}=\frac{\epsilon^{\mu\nu\rho}}{2\pi} K_{I,J} \partial_{\nu}a^J_{\rho}.
\end{equation}
Equation (\ref{eq: CS0_bulk}) predicts gapless modes at the boundary,
which is set to be at $y=0$.
The Lagrangian of the boundary modes is given by
\begin{equation}
\mathcal{L}_\mathrm{edge}=
\int\!\frac{dx}{4\pi}
(K_{I,J}\partial_t\phi_I \partial_x \phi_J
 -v_F\partial_x\phi_I \partial_x\phi_I ),
\end{equation}
with $\bm{\phi}:=(\phi^\uparrow,\phi^\downarrow)^T$
[$I,J=1,2$ and $(\uparrow,\downarrow)=(1,2)$].
Here $\phi^\uparrow$ and $\phi^\downarrow$ are the bosonic fields
describing gapless modes of up-spin and down-spin quasiparticles.
$v_F$ denotes the Fermi velocity.
The commutation relations of these fields are given by
\begin{equation}
[\phi_I(x),\phi_J(y)] =
-i\pi \rho^z_{I,J}\mathrm{sgn}(x-y) +i\pi\,\mathrm{sgn}(I-J).
\end{equation}
The transformation laws of these fields under $\hat{T},\hat{g},\hat{u}_\theta$
are given by Eq.\ (\ref{eq: trans-Tg_1})
with $(\phi^+,\phi^-)$ replaced by $(\phi^{\uparrow},\phi^{\downarrow})$.
Let us gauge the symmetry of fermion number parity.
The corresponding operator $\hat{P}_f$ transforms the bosonic fields as
\begin{subequations}
\begin{equation}
\hat{P}_f \bm{\phi} \hat{P}^{-1}_f= \bm{\phi} +\delta \bm{\phi}_{P_f},
\end{equation}
where
\begin{equation}
\delta \bm{\phi}_{P_f} =-\pi
\begin{pmatrix} 1 \\ 1 \end{pmatrix}.
\end{equation}
\end{subequations}
Gauging this symmetry is carried out by extending the action as follows:
\begin{enumerate}
\item[(I)]
A vortex (symmetry flux) is attached to quasiparticles
such that a quasiparticle with
$\bm{l}\in\mathbb{Z}^2$ going around a vortex
acquires a phase shift $\bm{l}\cdot\delta \bm{\phi}_{P_f}$
which is equal to the phase shift from the symmetry transformation
generated by $\hat{P}_f$.
This makes the $\bm{l}$ vectors of quasiparticles non-integer-valued.
\item[(II)]
The action (\ref{eq: CS0_bulk}) is extended so that the quasiparticles
with non-integer $\bm{l}$ vectors become elementary excitations.
\end{enumerate}
The symmetry flux introduced in the step (I) is
described by the vector
\begin{equation}
\bm{l}_{P_f}=
\begin{pmatrix}\frac{1}{2} \\ -\frac{1}{2}\end{pmatrix}.
\end{equation}
We can see this by calculating from Eq.~(\ref{eq: CS0_bulk})
the phase acquired by a quasiparticle with an $\bm{l}$ vector
going around the symmetry flux $\bm{l}_{P_f}$.
The phase is $2\pi\bm{l}^TK^{-1}\bm{l}_{P_f}$,
which should be compared with $\bm{l}\cdot\delta\bm{\phi}_{P_f}$.
With the flux contribution $\bm{l}_{P_f}$, quasiparticles are
effectively described by the following vectors:
\begin{subequations}
\label{eq: Mn}
\begin{equation}
\bm{l}':=M\bm{n},
\end{equation}
with 
\begin{equation}
M:=
\left(
\begin{array}{cc}
\frac{1}{2}  & -1 \\
-\frac{1}{2} & 0
\end{array}
\right),
\quad
\bm{n}\in \mathbb{Z}^2.
\end{equation}
\end{subequations}
In the step (II), we modify the action (\ref{eq: CS0_bulk})
so that all the quasiparticles are described by integer vectors,
which we identify with $\bm{n}$ in Eq.~(\ref{eq: Mn}).
The modified action then reads
\begin{subequations}
\begin{equation}
\mathcal{L}_{g}=
\int d^2\bm{x}\frac{\epsilon^{\mu\nu\rho}}{4\pi}{K_g}_{I,J}
\tilde{a}^I_\mu\partial_\nu \tilde{a}^J_\rho
 -n_Ij^\mu\tilde{a}^I_\mu,
\end{equation}
with 
\begin{equation}
K_g =
\left(
\begin{array}{cc}
4 &2  \\
2 &0
\end{array}
\right),
 \quad I=1,2,
\end{equation}
\end{subequations}
where fields $\tilde{a}^I_{\mu}=M_{J,I}a^J_\mu$
are the internal Chern-Simons gauge fields of this gauged system. 
The $K$ matrix for the gauged system gives
the mutual statistics of quasiparticles with integer vectors
($\bm{n},\bm{n}'\in\mathbb{Z}^2$) as
\begin{equation}
2\pi \bm{n}^T K^{-1}_{g} \bm{n}'
=
2\pi \bm{n}^T M^T K^{-1} M \bm{n}'.
\end{equation}
Without loss of generality, we can redefine the fields as
$\tilde{a}^I_\mu=X_{I,J}\tilde{a}'^J_\mu$ with $X\in GL(2,\mathbb{Z})$.
Choosing
\begin{equation}
X=
\left(
\begin{array}{cc}
1 & 0 \\
-1 & 1
\end{array}
\right),
\end{equation}
we obtain the following action of the QSHI$^{*}$ phase:
\begin{equation}
\mathcal{L}_{g}=
\int d^2\bm{x}\left(
\frac{\epsilon^{\mu\nu\rho}}{4\pi}
2\rho^x \tilde{a}'^{I}_{\mu} \partial_\nu \tilde{a}'^{J}_\rho
 -n'_Ij^{\mu}\tilde{a}'^{I}_{\mu}\right).
\end{equation}
This action predicts gapless edge modes at the boundary $y=0$,
which are described by the Lagrangian
\begin{subequations}
\label{eq: QSHIg_app}
\begin{equation}
\mathcal{L}_{g,\mathrm{edge}}=
\int \frac{dx}{4\pi} \left(
2\rho^x_{I,J} \partial_t\phi_I\partial_x\phi_J
-{V_g}_{I,J}\partial_x\phi_I\partial_x\phi_J \right),
\end{equation}
with 
\begin{equation}
\bm{\phi}^T=\left(\phi^s,-\phi^c\right).
\end{equation}
\end{subequations}
Here, $V_g$ is a positive-definite symmetric matrix.
This edge Lagrangian is equivalent to the Lagrangian in Eq.~(\ref{eq: QSHIg}).
The bosonic fields $\left(\phi^s,\phi^c\right)$ are related
to $\left(\phi^\uparrow,\phi^\downarrow\right)$ as follows:
\begin{eqnarray}
\left(
\begin{array}{cc}
\phi^s, & -\phi^c
\end{array}
\right)
&\leftrightarrow &
\left(
\begin{array}{cc}
\phi^\uparrow, & \phi^\downarrow
\end{array}
\right)
M(X^{T})^{-1},
\end{eqnarray}
or equivalently,
\begin{eqnarray}
\left(
\begin{array}{c}
2\phi^s  \\
2\phi^c
\end{array}
\right)
&\leftrightarrow &
\left(
\begin{array}{c}
\phi^\uparrow-\phi^\downarrow  \\
\phi^\uparrow+\phi^\downarrow  
\end{array}
\right).
\end{eqnarray}
From this correspondence, we obtain the transformation law of
fields $\phi^s$ and $\phi^c$ in Eq.\ (\ref{eq:Tg_QSHIg}).
It follows from the correspondence that
\begin{subequations}
\begin{eqnarray}
\hat{T}:\!e^{2i\phi^s}\!:\hat{T}^{-1}
&\leftrightarrow&
-:\!e^{-i(\phi^\uparrow-\phi^\downarrow)}\!: \,, \\
\hat{T}:\!e^{2i\phi^c}\!:\hat{T}^{-1}
&\leftrightarrow&
-:\!e^{i(\phi^\uparrow+\phi^\downarrow)}\!:.
\end{eqnarray}
\end{subequations}
One can see how the time-reversal operator transforms
the vertex operators in the pinning potentials
in Eq.~(\ref{eq: QSHIg_cos}) in the following way.
With the correspondence,
\begin{subequations}
\begin{eqnarray}
:\!e^{i(\phi^+-\phi^--2\phi^s)}\!:
&\leftrightarrow&
:\!e^{i(\phi^+-\phi^--\phi^\uparrow+\phi^\downarrow)}\!: \, , \\
:\!e^{i(\phi^++\phi^--2\phi^c)}\!:
&\leftrightarrow&
:\!e^{i(\phi^++\phi^--\phi^\uparrow-\phi^\downarrow)}\!: \, , 
\end{eqnarray}
\end{subequations}
applying the time-reversal operator yields
\begin{subequations}
\begin{eqnarray}
\hat{T}:\!e^{i(\phi^+-\phi^--2\phi^s)}\!:\hat{T}^{-1}
&\leftrightarrow&
:\!e^{i(\phi^+-\phi^--\phi^\uparrow+\phi^\downarrow)}\!: \, , \\
\hat{T}:\!e^{i(\phi^++\phi^--2\phi^c)}\!:\hat{T}^{-1}
&\leftrightarrow&
:\!e^{-i(\phi^++\phi^--\phi^\uparrow-\phi^\downarrow)}\!: \, ,
\qquad\quad
\end{eqnarray}
\end{subequations}
from which Eq.~(\ref{eq: QSHIg_cos}) follows.
Here the bosonic fields $\phi^+$ and $\phi^-$ describe the helical modes
of a QSHI.
In a similar way, we obtain Eq.~(\ref{eq: QSH*_v T}) by noting that
the vertex operators have following correspondence:
\begin{subequations}
\begin{eqnarray}
:\!e^{i\bm{v}_1\cdot\bm{\phi}}\!:
&\leftrightarrow&
:\!e^{i(\phi^+-\phi^{\uparrow})}\!: \,
=-i:\!e^{i\phi^+}\!\!:\,:\!e^{-i\phi^\uparrow}\!\!:\,, \\
:\!e^{i\bm{v}_2\cdot\bm{\phi}}\!:
&\leftrightarrow&
:\!e^{i(\phi^--\phi^{\downarrow})}\!: \,
=-i:\!e^{i\phi^-}\!\!:\,:\!e^{-i\phi^\downarrow}\!\!:\,,
\qquad
\end{eqnarray}
\end{subequations}
which implies the following relations:
\begin{subequations}
\begin{eqnarray}
\hat{T}:\!e^{i\bm{v}_1\cdot\bm{\phi}}\!:\hat{T}^{-1}
&\leftrightarrow&
-:\!e^{i(\phi^--\phi^{\downarrow})}\!:, \\
\hat{T}:\!e^{i\bm{v}_2\cdot\bm{\phi}}\!:\hat{T}^{-1}
&\leftrightarrow&
-:\!e^{i(\phi^+-\phi^{\uparrow})}\!: .
\end{eqnarray}
\end{subequations}


\begin{thebibliography}{50}%
\makeatletter
\providecommand \@ifxundefined [1]{%
 \@ifx{#1\undefined}
}%
\providecommand \@ifnum [1]{%
 \ifnum #1\expandafter \@firstoftwo
 \else \expandafter \@secondoftwo
 \fi
}%
\providecommand \@ifx [1]{%
 \ifx #1\expandafter \@firstoftwo
 \else \expandafter \@secondoftwo
 \fi
}%
\providecommand \natexlab [1]{#1}%
\providecommand \enquote  [1]{``#1''}%
\providecommand \bibnamefont  [1]{#1}%
\providecommand \bibfnamefont [1]{#1}%
\providecommand \citenamefont [1]{#1}%
\providecommand \href@noop [0]{\@secondoftwo}%
\providecommand \href [0]{\begingroup \@sanitize@url \@href}%
\providecommand \@href[1]{\@@startlink{#1}\@@href}%
\providecommand \@@href[1]{\endgroup#1\@@endlink}%
\providecommand \@sanitize@url [0]{\catcode `\\12\catcode `\$12\catcode
  `\&12\catcode `\#12\catcode `\^12\catcode `\_12\catcode `\%12\relax}%
\providecommand \@@startlink[1]{}%
\providecommand \@@endlink[0]{}%
\providecommand \url  [0]{\begingroup\@sanitize@url \@url }%
\providecommand \@url [1]{\endgroup\@href {#1}{\urlprefix }}%
\providecommand \urlprefix  [0]{URL }%
\providecommand \Eprint [0]{\href }%
\providecommand \doibase [0]{http://dx.doi.org/}%
\providecommand \selectlanguage [0]{\@gobble}%
\providecommand \bibinfo  [0]{\@secondoftwo}%
\providecommand \bibfield  [0]{\@secondoftwo}%
\providecommand \translation [1]{[#1]}%
\providecommand \BibitemOpen [0]{}%
\providecommand \bibitemStop [0]{}%
\providecommand \bibitemNoStop [0]{.\EOS\space}%
\providecommand \EOS [0]{\spacefactor3000\relax}%
\providecommand \BibitemShut  [1]{\csname bibitem#1\endcsname}%
\let\auto@bib@innerbib\@empty
\bibitem [{\citenamefont {Thouless}\ \emph {et~al.}(1982)\citenamefont
  {Thouless}, \citenamefont {Kohmoto}, \citenamefont {Nightingale},\ and\
  \citenamefont {den Nijs}}]{TKNN}%
  \BibitemOpen
  \bibfield  {author} {\bibinfo {author} {\bibfnamefont {D.~J.}\ \bibnamefont
  {Thouless}}, \bibinfo {author} {\bibfnamefont {M.}~\bibnamefont {Kohmoto}},
  \bibinfo {author} {\bibfnamefont {M.~P.}\ \bibnamefont {Nightingale}}, \ and\
  \bibinfo {author} {\bibfnamefont {M.}~\bibnamefont {den Nijs}},\ }\href
  {\doibase 10.1103/PhysRevLett.49.405} {\bibfield  {journal} {\bibinfo
  {journal} {Phys. Rev. Lett.}\ }\textbf {\bibinfo {volume} {49}},\ \bibinfo
  {pages} {405} (\bibinfo {year} {1982})}\BibitemShut {NoStop}%
\bibitem [{\citenamefont {Qi}\ \emph {et~al.}(2008)\citenamefont {Qi},
  \citenamefont {Hughes},\ and\ \citenamefont {Zhang}}]{Qi_3DTI2008}%
  \BibitemOpen
  \bibfield  {author} {\bibinfo {author} {\bibfnamefont {X.-L.}\ \bibnamefont
  {Qi}}, \bibinfo {author} {\bibfnamefont {T.~L.}\ \bibnamefont {Hughes}}, \
  and\ \bibinfo {author} {\bibfnamefont {S.-C.}\ \bibnamefont {Zhang}},\ }\href
  {\doibase 10.1103/PhysRevB.78.195424} {\bibfield  {journal} {\bibinfo
  {journal} {Phys. Rev. B}\ }\textbf {\bibinfo {volume} {78}},\ \bibinfo
  {pages} {195424} (\bibinfo {year} {2008})}\BibitemShut {NoStop}%
\bibitem [{\citenamefont {Schnyder}\ \emph {et~al.}(2008)\citenamefont
  {Schnyder}, \citenamefont {Ryu}, \citenamefont {Furusaki},\ and\
  \citenamefont {Ludwig}}]{Schnyder_classification_free_2008}%
  \BibitemOpen
  \bibfield  {author} {\bibinfo {author} {\bibfnamefont {A.~P.}\ \bibnamefont
  {Schnyder}}, \bibinfo {author} {\bibfnamefont {S.}~\bibnamefont {Ryu}},
  \bibinfo {author} {\bibfnamefont {A.}~\bibnamefont {Furusaki}}, \ and\
  \bibinfo {author} {\bibfnamefont {A.~W.~W.}\ \bibnamefont {Ludwig}},\ }\href
  {\doibase 10.1103/PhysRevB.78.195125} {\bibfield  {journal} {\bibinfo
  {journal} {Phys. Rev. B}\ }\textbf {\bibinfo {volume} {78}},\ \bibinfo
  {pages} {195125} (\bibinfo {year} {2008})}\BibitemShut {NoStop}%
\bibitem [{\citenamefont {Ryu}\ \emph {et~al.}(2010)\citenamefont {Ryu},
  \citenamefont {Schnyder}, \citenamefont {Furusaki},\ and\ \citenamefont
  {Ludwig}}]{Ryu_classification_free_2010}%
  \BibitemOpen
  \bibfield  {author} {\bibinfo {author} {\bibfnamefont {S.}~\bibnamefont
  {Ryu}}, \bibinfo {author} {\bibfnamefont {A.~P.}\ \bibnamefont {Schnyder}},
  \bibinfo {author} {\bibfnamefont {A.}~\bibnamefont {Furusaki}}, \ and\
  \bibinfo {author} {\bibfnamefont {A.~W.~W.}\ \bibnamefont {Ludwig}},\ }\href
  {http://stacks.iop.org/1367-2630/12/i=6/a=065010} {\bibfield  {journal}
  {\bibinfo  {journal} {New J. Phys.}\ }\textbf {\bibinfo {volume} {12}},\
  \bibinfo {pages} {065010} (\bibinfo {year} {2010})}\BibitemShut {NoStop}%
\bibitem [{\citenamefont {Kitaev}(2009)}]{Kitaev_classification_free_2009}%
  \BibitemOpen
  \bibfield  {author} {\bibinfo {author} {\bibfnamefont {A.}~\bibnamefont
  {Kitaev}},\ }\href {\doibase 10.1063/1.3149495} {\bibfield  {journal}
  {\bibinfo  {journal} {AIP Conf. Proc.}\ }\textbf {\bibinfo {volume} {1134}},\
  \bibinfo {pages} {22} (\bibinfo {year} {2009})}\BibitemShut {NoStop}%
\bibitem [{\citenamefont {Altland}\ and\ \citenamefont
  {Zirnbauer}(1997)}]{Altland}%
  \BibitemOpen
  \bibfield  {author} {\bibinfo {author} {\bibfnamefont {A.}~\bibnamefont
  {Altland}}\ and\ \bibinfo {author} {\bibfnamefont {M.~R.}\ \bibnamefont
  {Zirnbauer}},\ }\href {\doibase 10.1103/PhysRevB.55.1142} {\bibfield
  {journal} {\bibinfo  {journal} {Phys. Rev. B}\ }\textbf {\bibinfo {volume}
  {55}},\ \bibinfo {pages} {1142} (\bibinfo {year} {1997})}\BibitemShut
  {NoStop}%
\bibitem [{\citenamefont {Chen}\ \emph {et~al.}(2013)\citenamefont {Chen},
  \citenamefont {Gu}, \citenamefont {Liu},\ and\ \citenamefont
  {Wen}}]{chen_cohomology_3D}%
  \BibitemOpen
  \bibfield  {author} {\bibinfo {author} {\bibfnamefont {X.}~\bibnamefont
  {Chen}}, \bibinfo {author} {\bibfnamefont {Z.-C.}\ \bibnamefont {Gu}},
  \bibinfo {author} {\bibfnamefont {Z.-X.}\ \bibnamefont {Liu}}, \ and\
  \bibinfo {author} {\bibfnamefont {X.-G.}\ \bibnamefont {Wen}},\ }\href
  {\doibase 10.1103/PhysRevB.87.155114} {\bibfield  {journal} {\bibinfo
  {journal} {Phys. Rev. B}\ }\textbf {\bibinfo {volume} {87}},\ \bibinfo
  {pages} {155114} (\bibinfo {year} {2013})}\BibitemShut {NoStop}%
\bibitem [{\citenamefont {Chen}\ \emph {et~al.}(2011)\citenamefont {Chen},
  \citenamefont {Gu},\ and\ \citenamefont {Wen}}]{Chen_classification_1D_1}%
  \BibitemOpen
  \bibfield  {author} {\bibinfo {author} {\bibfnamefont {X.}~\bibnamefont
  {Chen}}, \bibinfo {author} {\bibfnamefont {Z.-C.}\ \bibnamefont {Gu}}, \ and\
  \bibinfo {author} {\bibfnamefont {X.-G.}\ \bibnamefont {Wen}},\ }\href
  {\doibase 10.1103/PhysRevB.83.035107} {\bibfield  {journal} {\bibinfo
  {journal} {Phys. Rev. B}\ }\textbf {\bibinfo {volume} {83}},\ \bibinfo
  {pages} {035107} (\bibinfo {year} {2011})}\BibitemShut {NoStop}%
\bibitem [{\citenamefont {Pollmann}\ \emph {et~al.}(2012)\citenamefont
  {Pollmann}, \citenamefont {Berg}, \citenamefont {Turner},\ and\ \citenamefont
  {Oshikawa}}]{Pollmann2012}%
  \BibitemOpen
  \bibfield  {author} {\bibinfo {author} {\bibfnamefont {F.}~\bibnamefont
  {Pollmann}}, \bibinfo {author} {\bibfnamefont {E.}~\bibnamefont {Berg}},
  \bibinfo {author} {\bibfnamefont {A.~M.}\ \bibnamefont {Turner}}, \ and\
  \bibinfo {author} {\bibfnamefont {M.}~\bibnamefont {Oshikawa}},\ }\href
  {\doibase 10.1103/PhysRevB.85.075125} {\bibfield  {journal} {\bibinfo
  {journal} {Phys. Rev. B}\ }\textbf {\bibinfo {volume} {85}},\ \bibinfo
  {pages} {075125} (\bibinfo {year} {2012})}\BibitemShut {NoStop}%
\bibitem [{\citenamefont {Senthil}\ and\ \citenamefont
  {Levin}(2013)}]{Senthil_IQHE_2013}%
  \BibitemOpen
  \bibfield  {author} {\bibinfo {author} {\bibfnamefont {T.}~\bibnamefont
  {Senthil}}\ and\ \bibinfo {author} {\bibfnamefont {M.}~\bibnamefont
  {Levin}},\ }\href {\doibase 10.1103/PhysRevLett.110.046801} {\bibfield
  {journal} {\bibinfo  {journal} {Phys. Rev. Lett.}\ }\textbf {\bibinfo
  {volume} {110}},\ \bibinfo {pages} {046801} (\bibinfo {year}
  {2013})}\BibitemShut {NoStop}%
\bibitem [{\citenamefont {Furukawa}\ and\ \citenamefont
  {Ueda}(2013)}]{furukawa_IQHE_2013}%
  \BibitemOpen
  \bibfield  {author} {\bibinfo {author} {\bibfnamefont {S.}~\bibnamefont
  {Furukawa}}\ and\ \bibinfo {author} {\bibfnamefont {M.}~\bibnamefont
  {Ueda}},\ }\href {\doibase 10.1103/PhysRevLett.111.090401} {\bibfield
  {journal} {\bibinfo  {journal} {Phys. Rev. Lett.}\ }\textbf {\bibinfo
  {volume} {111}},\ \bibinfo {pages} {090401} (\bibinfo {year}
  {2013})}\BibitemShut {NoStop}%
\bibitem [{\citenamefont {Takimoto}(2011)}]{Takimoto_2011}%
  \BibitemOpen
  \bibfield  {author} {\bibinfo {author} {\bibfnamefont {T.}~\bibnamefont
  {Takimoto}},\ }\href {\doibase 10.1143/JPSJ.80.123710} {\bibfield  {journal}
  {\bibinfo  {journal} {J. Phys. Soc. Jpn.}\ }\textbf {\bibinfo {volume}
  {80}},\ \bibinfo {pages} {123710} (\bibinfo {year} {2011})}\BibitemShut
  {NoStop}%
\bibitem [{\citenamefont {Dzero}\ \emph {et~al.}(2012)\citenamefont {Dzero},
  \citenamefont {Sun}, \citenamefont {Coleman},\ and\ \citenamefont
  {Galitski}}]{Dzero2012}%
  \BibitemOpen
  \bibfield  {author} {\bibinfo {author} {\bibfnamefont {M.}~\bibnamefont
  {Dzero}}, \bibinfo {author} {\bibfnamefont {K.}~\bibnamefont {Sun}}, \bibinfo
  {author} {\bibfnamefont {P.}~\bibnamefont {Coleman}}, \ and\ \bibinfo
  {author} {\bibfnamefont {V.}~\bibnamefont {Galitski}},\ }\href {\doibase
  10.1103/PhysRevB.85.045130} {\bibfield  {journal} {\bibinfo  {journal} {Phys.
  Rev. B}\ }\textbf {\bibinfo {volume} {85}},\ \bibinfo {pages} {045130}
  (\bibinfo {year} {2012})}\BibitemShut {NoStop}%
\bibitem [{\citenamefont {Zhang}\ \emph {et~al.}(2013)\citenamefont {Zhang},
  \citenamefont {Butch}, \citenamefont {Syers}, \citenamefont {Ziemak},
  \citenamefont {Greene},\ and\ \citenamefont {Paglione}}]{SmB6_Zhang2013}%
  \BibitemOpen
  \bibfield  {author} {\bibinfo {author} {\bibfnamefont {X.}~\bibnamefont
  {Zhang}}, \bibinfo {author} {\bibfnamefont {N.~P.}\ \bibnamefont {Butch}},
  \bibinfo {author} {\bibfnamefont {P.}~\bibnamefont {Syers}}, \bibinfo
  {author} {\bibfnamefont {S.}~\bibnamefont {Ziemak}}, \bibinfo {author}
  {\bibfnamefont {R.~L.}\ \bibnamefont {Greene}}, \ and\ \bibinfo {author}
  {\bibfnamefont {J.}~\bibnamefont {Paglione}},\ }\href {\doibase
  10.1103/PhysRevX.3.011011} {\bibfield  {journal} {\bibinfo  {journal} {Phys.
  Rev. X}\ }\textbf {\bibinfo {volume} {3}},\ \bibinfo {pages} {011011}
  (\bibinfo {year} {2013})}\BibitemShut {NoStop}%
\bibitem [{\citenamefont {Lu}\ \emph {et~al.}(2013)\citenamefont {Lu},
  \citenamefont {Zhao}, \citenamefont {Weng}, \citenamefont {Fang},\ and\
  \citenamefont {Dai}}]{SmB6_Lu2013}%
  \BibitemOpen
  \bibfield  {author} {\bibinfo {author} {\bibfnamefont {F.}~\bibnamefont
  {Lu}}, \bibinfo {author} {\bibfnamefont {J. Z.}~\bibnamefont {Zhao}}, \bibinfo
  {author} {\bibfnamefont {H.}~\bibnamefont {Weng}}, \bibinfo {author}
  {\bibfnamefont {Z.}~\bibnamefont {Fang}}, \ and\ \bibinfo {author}
  {\bibfnamefont {X.}~\bibnamefont {Dai}},\ }\href {\doibase
  10.1103/PhysRevLett.110.096401} {\bibfield  {journal} {\bibinfo  {journal}
  {Phys. Rev. Lett.}\ }\textbf {\bibinfo {volume} {110}},\ \bibinfo {pages}
  {096401} (\bibinfo {year} {2013})}\BibitemShut {NoStop}%
\bibitem [{\citenamefont {Fidkowski}\ and\ \citenamefont
  {Kitaev}(2010)}]{Z_to_Zn_Fidkowski_10}%
  \BibitemOpen
  \bibfield  {author} {\bibinfo {author} {\bibfnamefont {L.}~\bibnamefont
  {Fidkowski}}\ and\ \bibinfo {author} {\bibfnamefont {A.}~\bibnamefont
  {Kitaev}},\ }\href {\doibase 10.1103/PhysRevB.81.134509} {\bibfield
  {journal} {\bibinfo  {journal} {Phys. Rev. B}\ }\textbf {\bibinfo {volume}
  {81}},\ \bibinfo {pages} {134509} (\bibinfo {year} {2010})}\BibitemShut
  {NoStop}%
\bibitem [{\citenamefont {Ryu}\ and\ \citenamefont
  {Zhang}(2012)}]{RyuZhang2012}%
  \BibitemOpen
  \bibfield  {author} {\bibinfo {author} {\bibfnamefont {S.}~\bibnamefont
  {Ryu}}\ and\ \bibinfo {author} {\bibfnamefont {S.-C.}\ \bibnamefont
  {Zhang}},\ }\href {\doibase 10.1103/PhysRevB.85.245132} {\bibfield  {journal}
  {\bibinfo  {journal} {Phys. Rev. B}\ }\textbf {\bibinfo {volume} {85}},\
  \bibinfo {pages} {245132} (\bibinfo {year} {2012})}\BibitemShut {NoStop}%
\bibitem [{\citenamefont {Yao}\ and\ \citenamefont {Ryu}(2013)}]{YaoRyu2013}%
  \BibitemOpen
  \bibfield  {author} {\bibinfo {author} {\bibfnamefont {H.}~\bibnamefont
  {Yao}}\ and\ \bibinfo {author} {\bibfnamefont {S.}~\bibnamefont {Ryu}},\
  }\href {\doibase 10.1103/PhysRevB.88.064507} {\bibfield  {journal} {\bibinfo
  {journal} {Phys. Rev. B}\ }\textbf {\bibinfo {volume} {88}},\ \bibinfo
  {pages} {064507} (\bibinfo {year} {2013})}\BibitemShut {NoStop}%
\bibitem [{\citenamefont {Qi}(2013)}]{QiNJP2013}%
  \BibitemOpen
  \bibfield  {author} {\bibinfo {author} {\bibfnamefont {X.-L.}\ \bibnamefont
  {Qi}},\ }\href {http://stacks.iop.org/1367-2630/15/i=6/a=065002} {\bibfield
  {journal} {\bibinfo  {journal} {New J. Phys.}\ }\textbf {\bibinfo {volume}
  {15}},\ \bibinfo {pages} {065002} (\bibinfo {year} {2013})}\BibitemShut
  {NoStop}%
\bibitem [{\citenamefont {Gu}\ and\ \citenamefont {Levin}(2014)}]{GuLevin2014}%
  \BibitemOpen
  \bibfield  {author} {\bibinfo {author} {\bibfnamefont {Z.-C.}\ \bibnamefont
  {Gu}}\ and\ \bibinfo {author} {\bibfnamefont {M.}~\bibnamefont {Levin}},\
  }\href {\doibase 10.1103/PhysRevB.89.201113} {\bibfield  {journal} {\bibinfo
  {journal} {Phys. Rev. B}\ }\textbf {\bibinfo {volume} {89}},\ \bibinfo
  {pages} {201113} (\bibinfo {year} {2014})}\BibitemShut {NoStop}%
\bibitem [{\citenamefont {Fidkowski}\ \emph {et~al.}(2013)\citenamefont
  {Fidkowski}, \citenamefont {Chen},\ and\ \citenamefont
  {Vishwanath}}]{Fidkowski_Z162013}%
  \BibitemOpen
  \bibfield  {author} {\bibinfo {author} {\bibfnamefont {L.}~\bibnamefont
  {Fidkowski}}, \bibinfo {author} {\bibfnamefont {X.}~\bibnamefont {Chen}}, \
  and\ \bibinfo {author} {\bibfnamefont {A.}~\bibnamefont {Vishwanath}},\
  }\href {\doibase 10.1103/PhysRevX.3.041016} {\bibfield  {journal} {\bibinfo
  {journal} {Phys. Rev. X}\ }\textbf {\bibinfo {volume} {3}},\ \bibinfo {pages}
  {041016} (\bibinfo {year} {2013})}\BibitemShut {NoStop}%
\bibitem [{\citenamefont {Wang}\ \emph {et~al.}(2014)\citenamefont {Wang},
  \citenamefont {Potter},\ and\ \citenamefont
  {Senthil}}]{Wang_Potter_Senthil2014}%
  \BibitemOpen
  \bibfield  {author} {\bibinfo {author} {\bibfnamefont {C.}~\bibnamefont
  {Wang}}, \bibinfo {author} {\bibfnamefont {A.~C.}\ \bibnamefont {Potter}}, \
  and\ \bibinfo {author} {\bibfnamefont {T.}~\bibnamefont {Senthil}},\ }\href
  {\doibase 10.1126/science.1243326} {\bibfield  {journal} {\bibinfo  {journal}
  {Science}\ }\textbf {\bibinfo {volume} {343}},\ \bibinfo {pages} {629}
  (\bibinfo {year} {2014})}\BibitemShut {NoStop}%
\bibitem [{\citenamefont {Metlitski}\ \emph {et~al.}(2014)\citenamefont
  {Metlitski}, \citenamefont {Fidkowski}, \citenamefont {Chen},\ and\
  \citenamefont {Vishwanath}}]{Metlitski_3Dinteraction2014}%
  \BibitemOpen
  \bibfield  {author} {\bibinfo {author} {\bibfnamefont {M.~A.}\ \bibnamefont
  {Metlitski}}, \bibinfo {author} {\bibfnamefont {L.}~\bibnamefont
  {Fidkowski}}, \bibinfo {author} {\bibfnamefont {X.}~\bibnamefont {Chen}}, \
  and\ \bibinfo {author} {\bibfnamefont {A.}~\bibnamefont {Vishwanath}},\
  }\href@noop {} {\bibfield  {journal} {\bibinfo  {journal} {arXiv:1406.3032}\
  } (\bibinfo {year} {2014})}\BibitemShut {NoStop}%
\bibitem [{\citenamefont {You}\ and\ \citenamefont {Xu}(2014)}]{You_Cenke2014}%
  \BibitemOpen
  \bibfield  {author} {\bibinfo {author} {\bibfnamefont {Y.-Z.}\ \bibnamefont
  {You}}\ and\ \bibinfo {author} {\bibfnamefont {C.}~\bibnamefont {Xu}},\
  }\href {\doibase 10.1103/PhysRevB.90.245120} {\bibfield  {journal} {\bibinfo
  {journal} {Phys. Rev. B}\ }\textbf {\bibinfo {volume} {90}},\ \bibinfo
  {pages} {245120} (\bibinfo {year} {2014})}\BibitemShut {NoStop}%
\bibitem [{\citenamefont {Gu}\ and\ \citenamefont
  {Wen}(2014)}]{gu_supercohomology}%
  \BibitemOpen
  \bibfield  {author} {\bibinfo {author} {\bibfnamefont {Z.-C.}\ \bibnamefont
  {Gu}}\ and\ \bibinfo {author} {\bibfnamefont {X.-G.}\ \bibnamefont {Wen}},\
  }\href {\doibase 10.1103/PhysRevB.90.115141} {\bibfield  {journal} {\bibinfo
  {journal} {Phys. Rev. B}\ }\textbf {\bibinfo {volume} {90}},\ \bibinfo
  {pages} {115141} (\bibinfo {year} {2014})}\BibitemShut {NoStop}%
\bibitem [{\citenamefont
  {Kapustin}(2014{\natexlab{a}})}]{kapustin_bosonic_cobordisms2014_1}%
  \BibitemOpen
  \bibfield  {author} {\bibinfo {author} {\bibfnamefont {A.}~\bibnamefont
  {Kapustin}},\ }\href@noop {} {\bibfield  {journal} {\bibinfo  {journal}
  {arXiv:1403.1467}\ } (\bibinfo {year} {2014}{\natexlab{a}})}\BibitemShut
  {NoStop}%
\bibitem [{\citenamefont
  {Kapustin}(2014{\natexlab{b}})}]{kapustin_bosonic_cobordisms2014_2}%
  \BibitemOpen
  \bibfield  {author} {\bibinfo {author} {\bibfnamefont {A.}~\bibnamefont
  {Kapustin}},\ }\href@noop {} {\bibfield  {journal} {\bibinfo  {journal}
  {arXiv:1404.6659}\ } (\bibinfo {year} {2014}{\natexlab{b}})}\BibitemShut
  {NoStop}%
\bibitem [{\citenamefont {Kapustin}\ \emph {et~al.}(2014)\citenamefont
  {Kapustin}, \citenamefont {Thorngren}, \citenamefont {Turzillo},\ and\
  \citenamefont {Wang}}]{kapustin_fermionic_cobordisms2014}%
  \BibitemOpen
  \bibfield  {author} {\bibinfo {author} {\bibfnamefont {A.}~\bibnamefont
  {Kapustin}}, \bibinfo {author} {\bibfnamefont {R.}~\bibnamefont {Thorngren}},
  \bibinfo {author} {\bibfnamefont {A.}~\bibnamefont {Turzillo}}, \ and\
  \bibinfo {author} {\bibfnamefont {Z.}~\bibnamefont {Wang}},\ }\href@noop {}
  {\bibfield  {journal} {\bibinfo  {journal} {arXiv:1406.7329}\ } (\bibinfo
  {year} {2014})}\BibitemShut {NoStop}%
\bibitem [{\citenamefont {Lu}\ and\ \citenamefont
  {Vishwanath}(2012)}]{Lu_CS_2011}%
  \BibitemOpen
  \bibfield  {author} {\bibinfo {author} {\bibfnamefont {Y.-M.}\ \bibnamefont
  {Lu}}\ and\ \bibinfo {author} {\bibfnamefont {A.}~\bibnamefont
  {Vishwanath}},\ }\href {\doibase 10.1103/PhysRevB.86.125119} {\bibfield
  {journal} {\bibinfo  {journal} {Phys. Rev. B}\ }\textbf {\bibinfo {volume}
  {86}},\ \bibinfo {pages} {125119} (\bibinfo {year} {2012})}\BibitemShut
  {NoStop}%
\bibitem [{\citenamefont {Hsieh}\ \emph
  {et~al.}(2014{\natexlab{a}})\citenamefont {Hsieh}, \citenamefont {Morimoto},\
  and\ \citenamefont {Ryu}}]{Hsieh_CS_CPT_2014}%
  \BibitemOpen
  \bibfield  {author} {\bibinfo {author} {\bibfnamefont {C.-T.}\ \bibnamefont
  {Hsieh}}, \bibinfo {author} {\bibfnamefont {T.}~\bibnamefont {Morimoto}}, \
  and\ \bibinfo {author} {\bibfnamefont {S.}~\bibnamefont {Ryu}},\ }\href
  {\doibase 10.1103/PhysRevB.90.245111} {\bibfield  {journal} {\bibinfo
  {journal} {Phys. Rev. B}\ }\textbf {\bibinfo {volume} {90}},\ \bibinfo
  {pages} {245111} (\bibinfo {year} {2014}{\natexlab{a}})}\BibitemShut
  {NoStop}%
\bibitem [{\citenamefont {Wang}\ and\ \citenamefont
  {Senthil}(2014)}]{Wang_Senthil2014}%
  \BibitemOpen
  \bibfield  {author} {\bibinfo {author} {\bibfnamefont {C.}~\bibnamefont
  {Wang}}\ and\ \bibinfo {author} {\bibfnamefont {T.}~\bibnamefont {Senthil}},\
  }\href {\doibase 10.1103/PhysRevB.89.195124} {\bibfield  {journal} {\bibinfo
  {journal} {Phys. Rev. B}\ }\textbf {\bibinfo {volume} {89}},\ \bibinfo
  {pages} {195124} (\bibinfo {year} {2014})}\BibitemShut {NoStop}%
\bibitem [{\citenamefont {Fu}(2011)}]{Fu_TCI_2011}%
  \BibitemOpen
  \bibfield  {author} {\bibinfo {author} {\bibfnamefont {L.}~\bibnamefont
  {Fu}},\ }\href {\doibase 10.1103/PhysRevLett.106.106802} {\bibfield
  {journal} {\bibinfo  {journal} {Phys. Rev. Lett.}\ }\textbf {\bibinfo
  {volume} {106}},\ \bibinfo {pages} {106802} (\bibinfo {year}
  {2011})}\BibitemShut {NoStop}%
\bibitem [{\citenamefont {Tanaka}\ \emph {et~al.}(2012)\citenamefont {Tanaka},
  \citenamefont {Ren}, \citenamefont {Sato}, \citenamefont {Nakayama},
  \citenamefont {Souma}, \citenamefont {Takahashi}, \citenamefont {Segawa},\
  and\ \citenamefont {Ando}}]{Tanaka_TCI_SnTe2012}%
  \BibitemOpen
  \bibfield  {author} {\bibinfo {author} {\bibfnamefont {Y.}~\bibnamefont
  {Tanaka}}, \bibinfo {author} {\bibfnamefont {Z.}~\bibnamefont {Ren}},
  \bibinfo {author} {\bibfnamefont {T.}~\bibnamefont {Sato}}, \bibinfo {author}
  {\bibfnamefont {K.}~\bibnamefont {Nakayama}}, \bibinfo {author}
  {\bibfnamefont {S.}~\bibnamefont {Souma}}, \bibinfo {author} {\bibfnamefont
  {T.}~\bibnamefont {Takahashi}}, \bibinfo {author} {\bibfnamefont
  {K.}~\bibnamefont {Segawa}}, \ and\ \bibinfo {author} {\bibfnamefont
  {Y.}~\bibnamefont {Ando}},\ }\href@noop {} {\bibfield  {journal} {\bibinfo
  {journal} {Nat. Phys.}\ }\textbf {\bibinfo {volume} {8}},\ \bibinfo {pages}
  {800} (\bibinfo {year} {2012})}\BibitemShut {NoStop}%
\bibitem [{\citenamefont {Hsieh}\ \emph {et~al.}(2012)\citenamefont {Hsieh},
  \citenamefont {Lin}, \citenamefont {Liu}, \citenamefont {Duan}, \citenamefont
  {Bansil},\ and\ \citenamefont {Fu}}]{Hsieh_TCH_SnTe_2012}%
  \BibitemOpen
  \bibfield  {author} {\bibinfo {author} {\bibfnamefont {T.~H.}\ \bibnamefont
  {Hsieh}}, \bibinfo {author} {\bibfnamefont {H.}~\bibnamefont {Lin}}, \bibinfo
  {author} {\bibfnamefont {J.}~\bibnamefont {Liu}}, \bibinfo {author}
  {\bibfnamefont {W.}~\bibnamefont {Duan}}, \bibinfo {author} {\bibfnamefont
  {A.}~\bibnamefont {Bansil}}, \ and\ \bibinfo {author} {\bibfnamefont
  {L.}~\bibnamefont {Fu}},\ }\href@noop {} {\bibfield  {journal} {\bibinfo
  {journal} {Nat. Commun.}\ }\textbf {\bibinfo {volume} {3}},\ \bibinfo {pages}
  {982} (\bibinfo {year} {2012})}\BibitemShut {NoStop}%
\bibitem [{\citenamefont {Chiu}\ \emph {et~al.}(2013)\citenamefont {Chiu},
  \citenamefont {Yao},\ and\ \citenamefont
  {Ryu}}]{Chiu_TCI_classification2013}%
  \BibitemOpen
  \bibfield  {author} {\bibinfo {author} {\bibfnamefont {C.-K.}\ \bibnamefont
  {Chiu}}, \bibinfo {author} {\bibfnamefont {H.}~\bibnamefont {Yao}}, \ and\
  \bibinfo {author} {\bibfnamefont {S.}~\bibnamefont {Ryu}},\ }\href {\doibase
  10.1103/PhysRevB.88.075142} {\bibfield  {journal} {\bibinfo  {journal} {Phys.
  Rev. B}\ }\textbf {\bibinfo {volume} {88}},\ \bibinfo {pages} {075142}
  (\bibinfo {year} {2013})}\BibitemShut {NoStop}%
\bibitem [{\citenamefont {Morimoto}\ and\ \citenamefont
  {Furusaki}(2013)}]{Morimoto_classification2013}%
  \BibitemOpen
  \bibfield  {author} {\bibinfo {author} {\bibfnamefont {T.}~\bibnamefont
  {Morimoto}}\ and\ \bibinfo {author} {\bibfnamefont {A.}~\bibnamefont
  {Furusaki}},\ }\href {\doibase 10.1103/PhysRevB.88.125129} {\bibfield
  {journal} {\bibinfo  {journal} {Phys. Rev. B}\ }\textbf {\bibinfo {volume}
  {88}},\ \bibinfo {pages} {125129} (\bibinfo {year} {2013})}\BibitemShut
  {NoStop}%
\bibitem [{\citenamefont {Shiozaki}\ and\ \citenamefont
  {Sato}(2014)}]{shiozaki_classification_2014}%
  \BibitemOpen
  \bibfield  {author} {\bibinfo {author} {\bibfnamefont {K.}~\bibnamefont
  {Shiozaki}}\ and\ \bibinfo {author} {\bibfnamefont {M.}~\bibnamefont
  {Sato}},\ }\href {\doibase 10.1103/PhysRevB.90.165114} {\bibfield  {journal}
  {\bibinfo  {journal} {Phys. Rev. B}\ }\textbf {\bibinfo {volume} {90}},\
  \bibinfo {pages} {165114} (\bibinfo {year} {2014})}\BibitemShut {NoStop}%
\bibitem [{\citenamefont {Weng}\ \emph {et~al.}(2014)\citenamefont {Weng},
  \citenamefont {Zhao}, \citenamefont {Wang}, \citenamefont {Fang},\ and\
  \citenamefont {Dai}}]{Weng_Dai_TCIinYbB12_2014}%
  \BibitemOpen
  \bibfield  {author} {\bibinfo {author} {\bibfnamefont {H.}~\bibnamefont
  {Weng}}, \bibinfo {author} {\bibfnamefont {J.}~\bibnamefont {Zhao}}, \bibinfo
  {author} {\bibfnamefont {Z.}~\bibnamefont {Wang}}, \bibinfo {author}
  {\bibfnamefont {Z.}~\bibnamefont {Fang}}, \ and\ \bibinfo {author}
  {\bibfnamefont {X.}~\bibnamefont {Dai}},\ }\href {\doibase
  10.1103/PhysRevLett.112.016403} {\bibfield  {journal} {\bibinfo  {journal}
  {Phys. Rev. Lett.}\ }\textbf {\bibinfo {volume} {112}},\ \bibinfo {pages}
  {016403} (\bibinfo {year} {2014})}\BibitemShut {NoStop}%
\bibitem [{\citenamefont {Hsieh}\ \emph
  {et~al.}(2014{\natexlab{b}})\citenamefont {Hsieh}, \citenamefont {Liu},\ and\
  \citenamefont {Fu}}]{Hsieh_TCIinOxides2014}%
  \BibitemOpen
  \bibfield  {author} {\bibinfo {author} {\bibfnamefont {T.~H.}\ \bibnamefont
  {Hsieh}}, \bibinfo {author} {\bibfnamefont {J.}~\bibnamefont {Liu}}, \ and\
  \bibinfo {author} {\bibfnamefont {L.}~\bibnamefont {Fu}},\ }\href {\doibase
  10.1103/PhysRevB.90.081112} {\bibfield  {journal} {\bibinfo  {journal} {Phys.
  Rev. B}\ }\textbf {\bibinfo {volume} {90}},\ \bibinfo {pages} {081112}
  (\bibinfo {year} {2014}{\natexlab{b}})}\BibitemShut {NoStop}%
\bibitem [{\citenamefont {Isobe}\ and\ \citenamefont
  {Fu}(2015)}]{Isobe_Fu2015}%
  \BibitemOpen
  \bibfield  {author} {\bibinfo {author} {\bibfnamefont {H.}~\bibnamefont
  {Isobe}}\ and\ \bibinfo {author} {\bibfnamefont {L.}~\bibnamefont {Fu}},\
  }\href@noop {} {\bibfield  {journal} {\bibinfo  {journal} {arXiv preprint
  arXiv:1502.06962}\ } (\bibinfo {year} {2015})}\BibitemShut {NoStop}%
\bibitem [{\citenamefont {Liu}\ and\ \citenamefont {Fu}(2015)}]{SnTe_2D}%
  \BibitemOpen
  \bibfield  {author} {\bibinfo {author} {\bibfnamefont {J.}~\bibnamefont
  {Liu}}\ and\ \bibinfo {author} {\bibfnamefont {L.}~\bibnamefont {Fu}},\
  }\href {\doibase 10.1103/PhysRevB.91.081407} {\bibfield  {journal} {\bibinfo
  {journal} {Phys. Rev. B}\ }\textbf {\bibinfo {volume} {91}},\ \bibinfo
  {pages} {081407} (\bibinfo {year} {2015})}\BibitemShut {NoStop}%
\bibitem [{\citenamefont {Safaei}\ \emph {et~al.}(2015)\citenamefont {Safaei},
  \citenamefont {Galicka}, \citenamefont {Kacman},\ and\ \citenamefont
  {Buczko}}]{Safaei2015}%
  \BibitemOpen
  \bibfield  {author} {\bibinfo {author} {\bibfnamefont {S.}~\bibnamefont
  {Safaei}}, \bibinfo {author} {\bibfnamefont {M.}~\bibnamefont {Galicka}},
  \bibinfo {author} {\bibfnamefont {P.}~\bibnamefont {Kacman}}, \ and\ \bibinfo
  {author} {\bibfnamefont {R.}~\bibnamefont {Buczko}},\ }\href@noop {}
  {\bibfield  {journal} {\bibinfo  {journal} {ArXiv e-prints}\ } (\bibinfo
  {year} {2015})},\ \Eprint {http://arxiv.org/abs/1501.04728} {arXiv:1501.04728
  [cond-mat.mtrl-sci]} \BibitemShut {NoStop}%
\bibitem [{Note1()}]{Note1}%
  \BibitemOpen
  \bibinfo {note} {In two dimensions the group structure of the SPT phases with
  U(1)$\times Z_2$ symmetry is $\protect \mathbb {Z}\times \protect \mathbb
  {Z}_4$. In the noninteracting case each eigensector of $Z_2$ symmetry (having
  eigenvalues $+i$ and $-i$ for reflection) is insulators in class A and
  characterized by a (mirror) Chern number; the topological classification in
  the noninteracting case is $\protect \mathbb {Z}\times \protect \mathbb {Z}$.
  Under the assumption of the vanishing total Chern number (removing one
  $\protect \mathbb {Z}$ from $\protect \mathbb {Z}\times \protect \mathbb
  {Z}$), the classification is reduced from $\protect \mathbb {Z}$ to $\protect
  \mathbb {Z}_4$ by interactions, as shown in Ref.~\protect \rev@citealpnum
  {Isobe_Fu2015}}\BibitemShut {NoStop}%
\bibitem [{Note2()}]{Note2}%
  \BibitemOpen
  \bibinfo {note} {Without the time-reversal symmetry, the two eigensectors
  have independent Chern numbers, and the topological classification is
  $\protect \mathbb {Z}\times \protect \mathbb {Z}$}\BibitemShut {NoStop}%
\bibitem [{\citenamefont {Neupert}\ \emph {et~al.}(2011)\citenamefont
  {Neupert}, \citenamefont {Santos}, \citenamefont {Ryu}, \citenamefont
  {Chamon},\ and\ \citenamefont {Mudry}}]{Neupert_CS_2011}%
  \BibitemOpen
  \bibfield  {author} {\bibinfo {author} {\bibfnamefont {T.}~\bibnamefont
  {Neupert}}, \bibinfo {author} {\bibfnamefont {L.}~\bibnamefont {Santos}},
  \bibinfo {author} {\bibfnamefont {S.}~\bibnamefont {Ryu}}, \bibinfo {author}
  {\bibfnamefont {C.}~\bibnamefont {Chamon}}, \ and\ \bibinfo {author}
  {\bibfnamefont {C.}~\bibnamefont {Mudry}},\ }\href {\doibase
  10.1103/PhysRevB.84.165107} {\bibfield  {journal} {\bibinfo  {journal} {Phys.
  Rev. B}\ }\textbf {\bibinfo {volume} {84}},\ \bibinfo {pages} {165107}
  (\bibinfo {year} {2011})}\BibitemShut {NoStop}%
\bibitem [{\citenamefont {Levin}\ and\ \citenamefont
  {Stern}(2012)}]{Levin_CS_2012}%
  \BibitemOpen
  \bibfield  {author} {\bibinfo {author} {\bibfnamefont {M.}~\bibnamefont
  {Levin}}\ and\ \bibinfo {author} {\bibfnamefont {A.}~\bibnamefont {Stern}},\
  }\href {\doibase 10.1103/PhysRevB.86.115131} {\bibfield  {journal} {\bibinfo
  {journal} {Phys. Rev. B}\ }\textbf {\bibinfo {volume} {86}},\ \bibinfo
  {pages} {115131} (\bibinfo {year} {2012})}\BibitemShut {NoStop}%
\bibitem [{\citenamefont {Haldane}(1995)}]{haldane_nullvector}%
  \BibitemOpen
  \bibfield  {author} {\bibinfo {author} {\bibfnamefont {F. D. M}~\bibnamefont
  {Haldane}},\ }\href@noop {} {\bibfield  {journal} {\bibinfo  {journal} {Phys.
  Rev. Lett.}\ }\textbf {\bibinfo {volume} {74}},\ \bibinfo {pages} {2090}
  (\bibinfo {year} {1995})}\BibitemShut {NoStop}%
\bibitem [{Note3()}]{Note3}%
  \BibitemOpen
  \bibinfo {note} {In the classification sheme based on K theory, one finds
  that the relevant classifying space depends on whether one imposes additional
  reflection symmetry on class A (without time-reversal symmetry) or class AII
  (with time-reversal symmetry). In the former case the relevant classifying
  space is $C_0$ while it is $R_0$ in the latter case\cite
  {Morimoto_classification2013}}\BibitemShut {NoStop}%
\bibitem [{\citenamefont {Lu}\ and\ \citenamefont {Lee}(2014)}]{Lu_Lee2014}%
  \BibitemOpen
  \bibfield  {author} {\bibinfo {author} {\bibfnamefont {Y.-M.}\ \bibnamefont
  {Lu}}\ and\ \bibinfo {author} {\bibfnamefont {D.-H.}\ \bibnamefont {Lee}},\
  }\href {\doibase 10.1103/PhysRevB.89.205117} {\bibfield  {journal} {\bibinfo
  {journal} {Phys. Rev. B}\ }\textbf {\bibinfo {volume} {89}},\ \bibinfo
  {pages} {205117} (\bibinfo {year} {2014})}\BibitemShut {NoStop}%
\bibitem [{\citenamefont {Lu}\ and\ \citenamefont
  {Vishwanath}(2013)}]{LuVishwanath_gauge2013}%
  \BibitemOpen
  \bibfield  {author} {\bibinfo {author} {\bibfnamefont {Y.-M.}\ \bibnamefont
  {Lu}}\ and\ \bibinfo {author} {\bibfnamefont {A.}~\bibnamefont
  {Vishwanath}},\ }\href@noop {} {\bibfield  {journal} {\bibinfo  {journal}
  {arXiv preprint arXiv:1302.2634}\ } (\bibinfo {year} {2013})}\BibitemShut
  {NoStop}%
\end{thebibliography}
%

%
%
%
\end{document}